\documentclass[manuscript]{aastex}

\usepackage{amsmath}
\usepackage{diagbox}
\usepackage{lscape}
\shorttitle{Deconvolution Computation Algorithm}
\shortauthors{Li Guangwei et al.}

\begin{document}


\title{A Practical Deconvolution Computation Algorithm to Extract 1D Spectra from 2D Images of Optical Fiber Spectroscopy}


\author{Li Guangwei, Zhang Haotong and Bai Zhongrui}
\affil{Key Lab for Optical Astronomy, National Astronomical Observatories, Chinese Academy of
Sciences, Beijing 100012}
\email{lgw@bao.ac.cn}



\begin{abstract}
\citet{bol10} presented a promising deconvolution method to extract one-dimensional (1D) spectra from a  two-dimensional (2D) optical fiber spectral CCD (Charge-Coupled Device) image. The method could eliminate the PSF (Point-Spread Function) difference between fibers, extract spectra to the photo noise level as well as improve the resolution. But the method is limited by its huge computation requirement and thus can not be implemented in actual data reduction. In this paper, we develop a practical computation method to solve the computation problem.
 The new computation method can deconvolve a 2D fiber spectral image of any size with actual PSFs, which may vary with positions. Our method does not require large memory and can extract a 4k $\times$ 4k noise free CCD image with 250 fibers in 2 hours. To make our method more practical,
 we further consider the influence of noise, which is thought to be an intrinsic ill-posed problem in  deconvolution  algorithms. We modify our method with a Tikhonov regularization
  item to depress the method induced noise, we do a series of simulations  to test how our method performs under more
  real situations with poisson noise and extreme cross talk. Compared with the results of traditional extraction methods,  i.e., the Aperture Extraction Method and the Profile Fitting Method, our method has the least residual and   influence by cross talk. 
  For the noise-added image, the computation speed  does not depend very much on fiber distance,  the SNR converges  in 2 to 4 iterations, the computation times are about 3.5 hours for the extreme fiber distance and about 2 hours for  non-extreme cases. A better balance between the computation time and result precision could be achieved by setting the
  precision threshold similar  to the noise level.
  We finally apply our method to real LAMOST (Large sky Area Multi-Object fiber Spectroscopic Telescope, a.k.a. Guo Shou Jing  Telescope) data. We find  that the 1D spectrum extracted
  by our method has both higher SNR and resolution than  the traditional methods, but there are still some suspicious weak features  possibly caused by the method around the strong emission lines. As we have demonstrated, our deconvolution method  has solved  the computation
  problem and progressed in dealing with the noise influence.  Multi-fiber spectra extracted by our method  will have higher resolution and signal to noise ratio thus will provide more accurate information (such as higher radial velocity and metallicity  measurement accuracy in stellar physics) to astronomers than traditional methods.

\end{abstract}

\keywords{Algorithm: general --- deconvolution, optimization, spectrum, optical fiber: individual --- (LAMOST)}

\section{Introduction}
To extract 1D spectra from 2D CCD images of optical fiber spectrographs,
there are 3 extraction methods that we often use.
\par
The first method is the Aperture Extraction Method (AEM, hereafter).
This method sums up counts in CCD pixels within a given aperture along a fiber trace on
both sides of its trace center. It is quick and easy but fails to deal with
cross talk between fibers when the distance is close. Meanwhile, simple
summation will give higher weights to the pixels with lower Signal to Noise Ratio
(SNR) at the wing of the profile, which will deteriorate the whole extraction.
\par

To solve this problem, \cite{hor86} and \cite{rob86} introduced independently
a method named the optimal extraction method. This method gives
higher weights to pixels with higher SNR. After that, many people kept improving
this method, for example, \cite{leb10} applied the method to extract spectra
with distorted traces. \cite{muk90}, \cite{mar89}, \cite{ver90}, \cite{hyn02} and
\cite{pis02}  developed many techniques to extract multi-fiber CCD images.
These technics are successful in some cases, yet they can not work correctly
when there is cross talk between fibers. Because noise of current CCD is very low,
usually several counts per pixel, according to equation 11 in \cite{hor86}, the
optimal extraction method is formally the same as AEM.
\par

With the development of the optimal extraction method, the Profile Fitting
Method (PFM, here after) arises. This method tries to simultaneously  fit CCD counts of different fibers along the spatial
direction with  bell-like profile functions at each wavelength. The spectral fluxes can be
derived from fitting the parameters of the profile function. Thus, one can solve the
cross talk problem by fitting the overlap  between fibers, e.g., \cite{bal00}, \cite{kel06}, \cite{san06} and \cite{sha10}.
\par
\begin{figure}[h]
  \begin{minipage}[t]{0.2\linewidth}
  \centering
   \includegraphics[width=22mm,height=58mm]{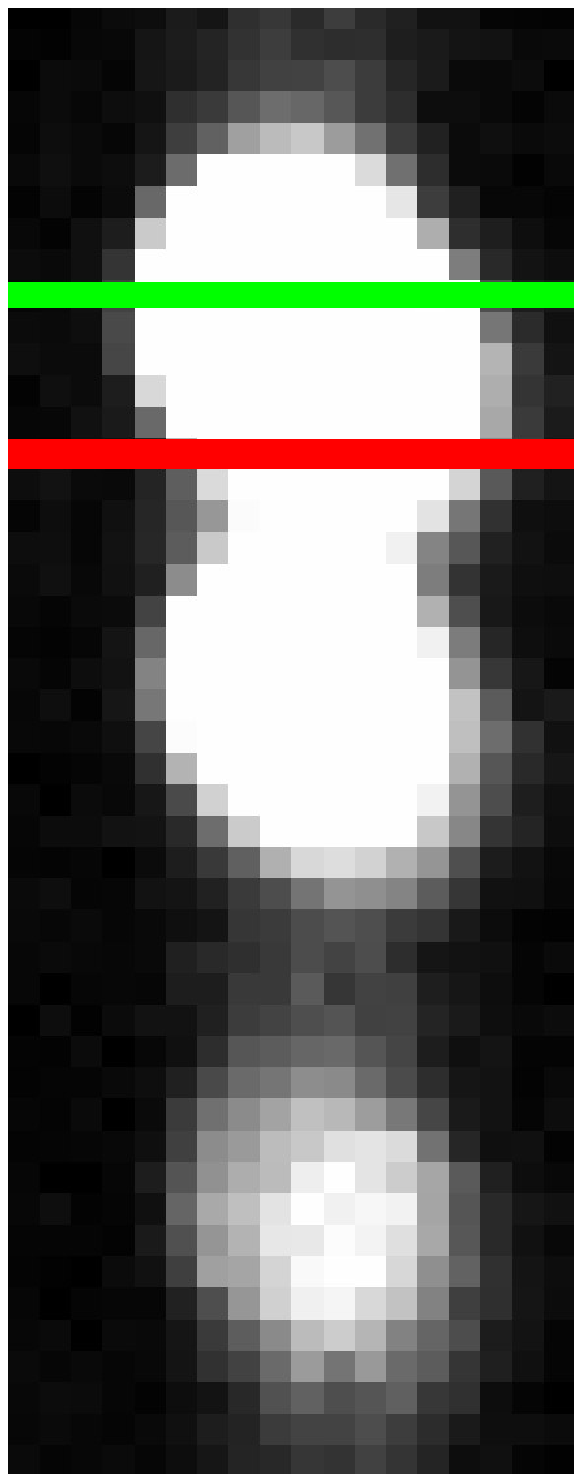}

  \end{minipage}%
  \begin{minipage}[t]{0.495\textwidth}
  \centering
   \includegraphics[width=100mm,height=58mm]{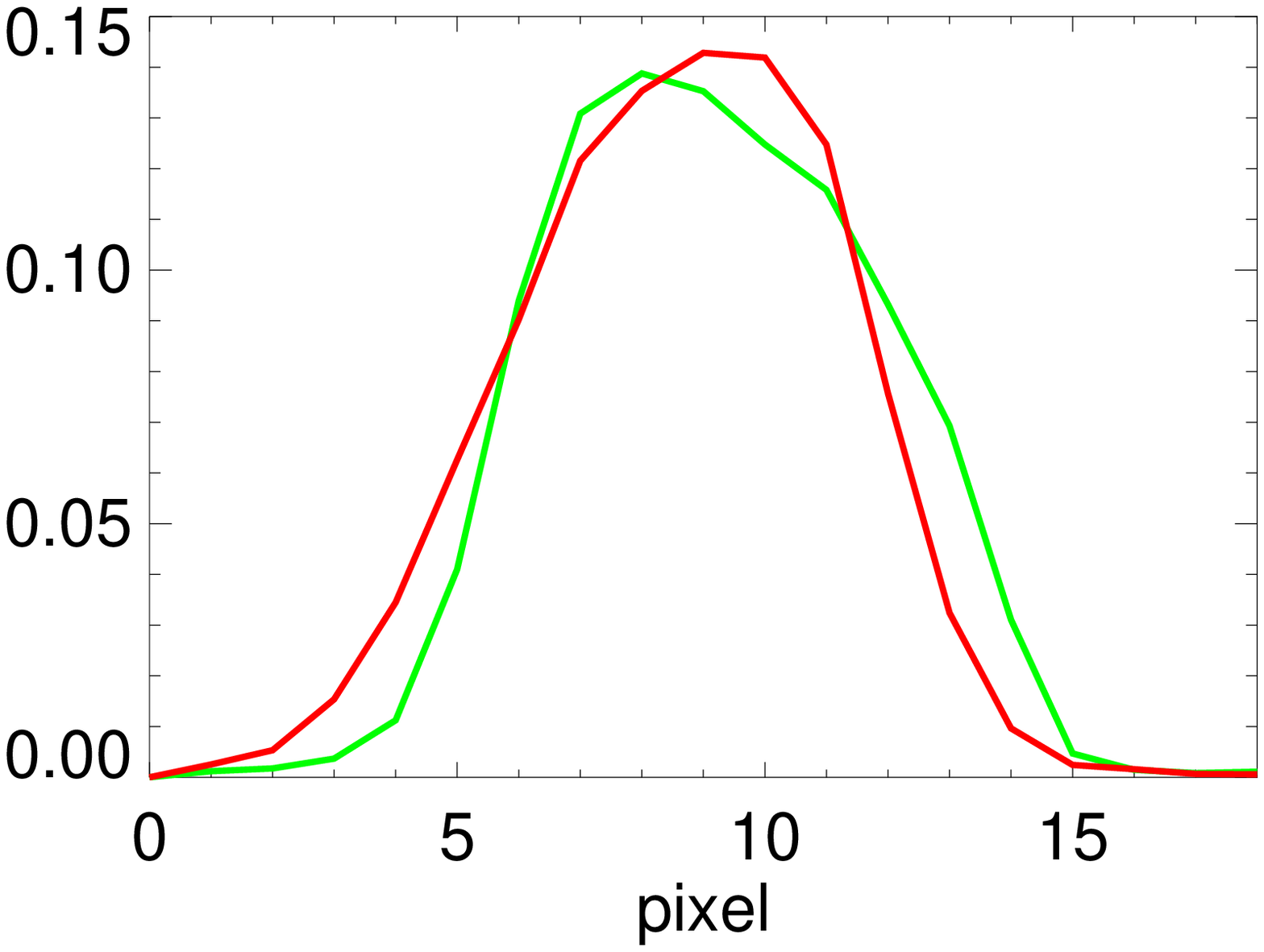}

  \end{minipage}%
  \caption{The left panel is a piece of arc lamp fiber spectral image of LAMOST (\cite{cui12}),
   bright spots are emission lines. The green and the red lines indicate rows where we
  sample the 1D profiles on the same arc emission line. The right panel illustrates
  normalized profiles  sampled on the row with the same colors in the left. The profile difference is obvious.
   We can see that extraction method based on $I(x)\times I(y)$ can not work properly
  when the profile is position dependent. }

  \label{psf-1D}
\end{figure}

These traditional extraction methods all extract flux row by row, which assumes
the flux distribution at every wavelength in a 2D CCD image is only a function of
the spatial direction and independent of the 2D
PSF (see \cite{bol10}). That is, flux $I(x,y)$  measured at pixel $(x, y)$
can be written as: $$I(x,y) = I(x)\times I(y),$$ this equation indicates that the profiles
along $y$ direction at different $x$ are not correlated, so that we can use the
same 1D profile to extract fluxes at different wavelengths. Considering the actual
optical aberration, this requirement is too rigorous (see Fig. \ref{psf-1D}).

The fourth method is the deconvolution method, which was introduced by \cite{bol10}.
The principle of this method is that if we take a 2D spectral image as the convolution of PSFs
and 1D spectra, when  the PSFs at all positions  are known, the 1D spectra can be obtained
by deconvolution. This method can eliminate the profile difference between fibers, caused by
the instrument optical aberration. However, the biggest drawback of the deconvolution method,
as discussed by \cite{bol10}, is that it needs a calibration matrix made from PSFs,
which is too huge to be stored in memory of common computers, leaving alone computation.
In this paper, we will give a new solution scheme to solve this problem. We also try to apply the
deconvolution method to simulations with noise and find that the method is very sensitive to noise,
that is, the solution is very unstable when noise is induced. In this paper, we also describe how
to partly solve the noise problem by adding a Tikhonov item in our method.

\par
This paper is organized as following: In Section 2, we introduce how to construct the PSFs
from the arc lamp lines. Then in Section 3, our objective functions and algorithm are discussed.
Computation tests on simulations with noise and serious cross talk as well as real telescope data will be illustrated in Section 4.
The last section is our conclusion and discussion.
\par

\section{Constructing PSFs}  \label{sect2}
Due to instrument aberration, PSFs at different positions on a CCD image are different.
To fully sample the actual PSFs on the CCD chip, non-blended emission lines at all positions
on the CCD is necessary. This condition is hard to be satisfied without an optical frequency comb.
Instead, we can sample basic PSFs from the arc lamp emission lines which are sparsely distributed
along the wavelength direction. Since the PSFs change smoothly with position, for positions where there is
no emission line, the PSF can be linearly interpolated from those basic PSFs.
\par

 We select single lines from an arc lamp image. These arc lines should have high SNR
and the distances between them should be as short as possible. Weak arc lines can be improved with
another longer exposure. Then we calculate the centroid $(x,y)$ of each good arc line, where $y$
is the coordinate in the spatial direction and $x$ is  in the wavelength direction.
The PSF of each arc line is normalized and then fitted with a uniform B-spline surface. Then a basic PSF is gotten.
\par
Assuming $PSF_0(x_0,y_0)$ is the PSF with center $(x_0, y_0)$, $PSF_1(x_1,y_1)$ and
$PSF_2(x_2,y_2)$ are two closest basic PSF neighbors on different sides of $PSF_0$ at $(x_1, y_1)$ and $(x_2, y_2)$ respectively.
To calculate $PSF_0$ from $PSF_1$ and $PSF_2$, we firstly move $PSF_1$ from $(x_1, y_1)$ to
$(x_1, y_0)$, and $PSF_2$ from $(x_2, y_2)$ to $(x_2, y_0)$, then get 2 new PSFs: $PSF'_1$
and $PSF'_2$. If $PSF[i,j]$  denotes the PSF in the $i$th row and the $j$th column on the CCD image,
then we can get $PSF_0$ by the formula:

\begin{equation} \label{psf_inter}
PSF_0[i,j] = \frac{x_2-x_0}{x_2-x_1}PSF'_1[i,j] + \frac{x_0-x_1}{x_2-x_1}PSF'_2[i,j].
\end{equation}

As for the PSFs close to the image edge, we can extrapolate or replicate the PSF of the closet arc line.
For the purpose of this paper, it's not necessary to use complicated algorithm, so the later is adopted.
To save compuation memory, only the PSFs of arc lines are stored as basic PSFs during the program running,
other PSFs are calculated by equation \ref{psf_inter} whenever necessary.
\par
In some instruments, distances of neighboring fibers may be so close that PSFs  can not be measured
simultaneously in a single arc exposure due to fiber-to-fiber cross talk. This could be solved by masking all the
other fibers on the focal plane while taking an arc exposure of a single fiber or fibers separated by  large distances
on the CCD chip at the same time. Combining PSFs from different exposures will give a full sample of PSFs on the CCD chip.
However, the algorithm should not be applied if masking  neighboring fibers is not possible for instruments
with significant fiber cross talk.

\section{The Objective Function and The Deconvolution Algorithm}

\subsection{The Objective Function}

Convolved with  profiles of fiber and  instrument, a monochromatic light beam
entering a spectrograph will end up with a bell shape PSF rather than a
$\delta$ function on the CCD. A 2D fiber spectrum can be treated as a lineup
of these single PSFs at different wavelengths with different intensities:

\begin{equation}
\label{objfuc01}
S(\lambda, y) = \int^{+\infty}_{-\infty} C(x) PSF_{\lambda, y}(\lambda - x, y) dx,
\end{equation}
where $S(\lambda, y)$ is the count recorded by the CCD at wavelength $\lambda$
and spatial coordinate $y$, and $C(x)$ is the 1D spectrum we want to derive.

\par

Considering cross talk between fibers in a multi-fiber spectral image,
the final count $F(\lambda, y)$ at position $(\lambda,y)$ can be written as:

\begin{equation} \label{objfucanal}
F(\lambda, y) = \sum^{N_f}_{i=1} S_{i}(\lambda, y ) + \epsilon_{\lambda, y},
\end{equation}
where $S_{i}(\lambda, y )$ is the count of the $i$th spectrum at wavelength
$\lambda$ and spatial coordination $y$, $N_f$ is the number of fibers, and $\epsilon_{\lambda, y}$ is noise.

\par
If we have an image of $N$ pixels in the dispersion direction and $M$ pixels in
the spatial direction, then for an actual computer calculation, equation \ref{objfucanal}
can be written in a discrete form:

\begin{equation} \label{objfunc}
F[k,m] = \sum^{N_f}_{i=1}\sum^{N}_{j=1} c_{i,j}PSF_{i,j}[k,m] + \varepsilon[k,m],
\end{equation}
where $F[k,m]$ is the flux at the $k$th row and $m$th column of the CCD, $c_{i,j}$
is the flux recorded at the $i$th fiber and the $j$th row   and $\varepsilon[k,m]$
is the noise at the $k$th row and $m$th column.
\par
From equation \ref{objfunc}, we can see that it is a least-quare problem.
We rewrite  equation \ref{objfunc} in  matrix form:

\begin{equation} \label{objfunc_matrix}
F = AC + \varepsilon,
\end{equation}
where $A$ is the calibration matrix with $PSF_{i,j}[k,m]$ as its element at
the $(k*M + m)$th row and the $(i*N + j)$th column, and $C$ is a vector with
$c_{i,j}$ as its $(i*N + j)$th element. For convenience of description, we
always assume that noise $\varepsilon$ is independent and identically distributed
in this paper,  other cases can be easily changed into this case by a
linear transformation.

\par

For gaussian noise, the objective function of   equation \ref{objfunc} is:

\begin{equation}\label{obj}
 \min_{c_{i,j}} \sum^{N}_{k=1} \sum^{M}_{m=1}(F[k,m] -
\sum^{N_f}_{i=1} \sum^{N}_{j=1} c_{i,j}PSF_{i,j}[k,m])^2 .
\end{equation}

For poisson noise, the objective function of   equation \ref{objfunc} is:

\begin{equation}\label{obj_poisson}
 \min_{c_{i,j}} \sum^{N}_{k=1} \sum^{M}_{m=1}\frac{(F[k,m] -
\sum^{N_f}_{i=1} \sum^{N}_{j=1}c_{i,j}PSF_{i,j}[k,m])^2}{F[k,m]} .
\end{equation}

We rewrite function \ref{obj} in matrix form:
\begin{equation} \label{Ge}
\min_C (F-AC)^T(F-AC),
\end{equation}

and its solution is
\begin{equation} \label{Guassian_solution}
C = (A^{T}A)^{-1}A^TF.
\end{equation}

If we denote $W$ as a diagonal matrix with $\frac{1}{F[k,m]}$ as its $(k*M + m)$th
diagonal element, then we can rewrite function \ref{obj_poisson} as
\begin{equation} \label{Pe}
\min_C (F-AC)^TW(F-AC),
\end{equation}
and its solution is
\begin{equation} \label{Poisson_solution}
C = (A^{T}WA)^{-1}A^TWF.
\end{equation}

In the case of  LAMOST (\cite{cui12}),  i.e., 250  spectra on a 4k$\times$4k
image, if we directly solve equation \ref{Guassian_solution} or
\ref{Poisson_solution} as in the paper of \cite{bol10}, the size of the
calibration matrix $A$ will be  $(4k \times 250) \times (4k \times 4k) = 1.6 \times10^{13}$,
so a double precision will require memory about 128TBytes.
Even though matrix $A$ is a sparse matrix, it can not be easily stored or computed.
Therefore, we can not solve this problem directly by the least-square method.
However, from function \ref{obj} and \ref{obj_poisson} we can see, they are
both optimization problems. Therefore, we try to solve it by an alternative
optimization method.
\par

\subsection{Algorithm}\label{sec:Algor}

For concision of expression, we will only discuss function \ref{Ge},
but the following procedure is also suitable for function \ref{Pe}.
\par
As discussed above, direct computation of $C$ from equation
\ref{Guassian_solution} will require unacceptable hardware resources.
But if we begin with an initial guess of $C_0$, then change $C_0$ in
a way that could reduce the value of function \ref{Ge} gradually,
 the final solution $C$ will be found when the function converges
to its global minimum. To further reduce the computation requirement,
every time we only change a small block of $C$ to minimize the value
of function \ref{Ge} while leaving other elements fixed. Then the next
block of $C$ will be changed to get another local minimum. At last,
the global solution will be approached when the value of
function \ref{Ge} stops to decrease or the decrease  is less than a
given precision.

\par
To begin the iteration process, we first chose an initial guess,
$C_0$, and a small piece of element block in $C_0$, for example,
the first $k$ elements ($k$ is a small integer). All the elements
  in this block will be changed to minimize objective function
\ref{Ge}. Next, the $\lceil k/2 \rceil +1$th$, ..., \lceil k/2 \rceil +k$th
element of the new $C_0$ will be selected as the new block,  elements
inside the block will be modified to minimize the objective function while
keeping other elements unchanged. Each time the block is shifted forward
by $\lceil k/2 \rceil$. When all elements in $C_0$ have been updated,
the new $C_0$ will be taken as the input in the next iteration.
\par
Since all PSFs in calibration matrix A are known,  objective function
\ref{Ge} has only one global minimum, which is equation \ref{Guassian_solution}.
If we keep repeating the above process, the objective function value will keep
decreasing and  achieve its global minimum point at last.
\par

We denote
$$R := F-AC_0 $$ and  $$res := R^{T}R.$$

For a given $C_0$ and a block made up of elements from the $(s+1)$th to the $(s+k)$th,
let $\delta C := (0,...,0,x_{s+1},...,x_{s+k},0,...,0)^T$, $X_s := (x_{s+1},...,x_{s+k})^T$
and define the $(s+1)$th, $..., (s+k)$ th columns in $A$ as submatrix $\bar{A}$,
then $A\delta C \equiv \bar{A}X_{s}$. If we change $C_0$ by $\delta C$, the objective function becomes
\begin{flalign}\label{der_proc}
\begin{split}
            & (F-A(C_0 + \delta C))^T(F-A(C_0+\delta C))\\
             =& (R - A\delta C)^T(R - A\delta C)\\
             = &  (R - \bar{A}X_{s})^T(R -\bar{A}X_{s})\\
             =  & res - 2R^T\bar{A}X_{s}+ X_{s}^T\bar{A}^T\bar{A}X_{s}.
\end{split}
\end{flalign}

If we let $X_s = (\bar{A}^T\bar{A})^{-1}\bar{A}^TR$, where the objective function
achieves its local minimum in the block $X_s$, then the value of the objective
function becomes
$$res - (R^T\bar{A})(\bar{A}^T\bar{A})^{-1}(R^T\bar{A})^T \leq res.$$
The second term in the left side of the formula is a non-negative quadratic term, this means the objective function will decrease when we change
$C$ locally by $\delta C$. If we keep iterating in deferent blocks,
the objective function will finally converge to its global minimum.

\par

The algorithm can be described as following:

Step 0: Given a value $\epsilon > 0$, the size of block: $k$, and the
size of variable $C$: $N$, $0 < k \ll N$;
\par
Step 1: Let $C_0 := 0$, $s := 0$, $r_1 := 0$. For function \ref{Ge}, let $R := F$ and $res := R^TR$;
\par
Step 2: If $s + k > N$, $s := N$, goto setp 3; else let $\bar{A}$ be
the submatrix consisting of the $s+1$th $,...,s+k$th columns in $A$.
We denote $X_s := (\bar{A}^T\bar{A})^{-1}\bar{A}^TR$. Let $r_1 := r_1 +
2R^T\bar{A}X_{s} - X_{s}^T\bar{A}^T\bar{A}X_{s}$ ;
$(c_{s + 1},...,c_{s + k})^T := (c_{s + 1},...,c_{s + k})^T + X_s$;
$R := R - \bar{A}X_{s}$ and $s := s + \lceil k/2 \rceil + 1$; goto step 2;
\par
Step 3: If $s - k < 0$, $s := 0$, then go to step 4; else let $\bar{A}$
be the submatrix consisting of the $s - k + 1$th $,...,s$th columns in
$A$. We denote $X_s := (\bar{A}^T\bar{A})^{-1}\bar{A}^TR$. Let $r_1 :=  r_1 +
2R^T\bar{A}X_{s} - X_{s}^T\bar{A}^T\bar{A}X_{s}$ ;
$(c_{s - k + 1},...,c_{s})^T := (c_{s - k + 1},...,c_{s})^T + X_s$;
$R := R - \bar{A}X_{s}$ and $s := s - \lceil k/2 \rceil - 1$;

\par
Step 4: $res := res - r_1$; If $r_1 < \epsilon$, stop; else let $r_1=0$, goto step 2;
\par

We call this algorithm   \emph{the Direct Deconvolution Algorithm}, or DDA, hereafter.
\par
Since $k$ is small, $\bar{A}X_{s}$ in the algorithm can be sparsely
stored and calculated. In principle, the algorithm can deal with an
image of any size with the PSFs dependent on positions.
\par
In addition, for a multi-processor computer, if we use one processor
to calculate a single fiber at a time, this algorithm can be easily transferred
into a parallel algorithm. For example, the size of our constructed LAMOST  PSFs
 is 15 pixels $\times$ 15 pixels, if the distances between fibers
are all greater than 8 pixels, then the odd fibers do not affect each other,
neither do the even fibers. Then the following computation scheme could
be applied:  we first use processors to compute  the odd fibers
in parallel,  after all the odd fibers have been computed, the
even fibers will be computed in parallel. The above parallel computation
process will be iterated until the solution is found. In general, if  the minimal
distance between the fibers is $D$,  we can divide all the fibers into
$\lceil \frac{15}{D} \rceil$ groups. We denote
$\lceil \frac{15}{D} \rceil \triangleq G$, and $i=1, 2, ..., 250$
as the sequence numbers of the 250 fibers. The $g$th group consists of fibers with
$i\  mod \ G = g$, where $g = 0, 1, 2, ..., G-1$. Thus, fibers in the
same group have no cross talk,  our parallel algorithm can be applied
in every group.

\section{Computation tests}
\label{sect:Obs}
Throughout this paper, the computer to test our algorithm is a
MacBook Pro with OS Windows XP, Intel (R) Core (TM) i7, 2.67GHz
CPU and 2G memory. The programming language is VC++ 6.0.
\subsection{Applied to a simulated 2D image based on LAMOST data}
\label{sim_data}

\begin{figure}
\plotone{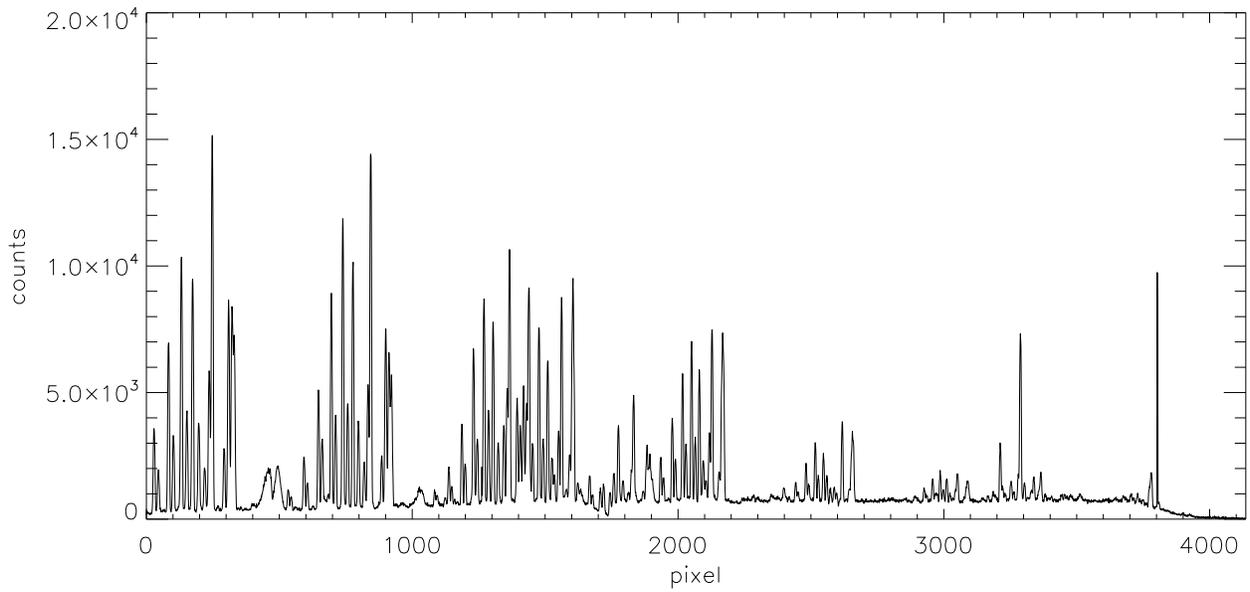}
\caption{One of the input 1D spectra used to generate the 2D simulation image.
It is the red part of a LAMOST spectrum. As we can see, there are many
strong OH sky emission lines. These lines will help us to evaluate how our
deconvolution method works on reliability, SNR and resolution.
\label{inputspec}}
\end{figure}

To test our algorithm, we have constructed a simulation 2D image
based on LAMOST data. 33 emission lines per fiber from a LAMOST arc
image are selected as basic PSFs to construct a $4K \times 4K$
CCD image with 250 fibers. 250 input 1D spectra are  extracted
from a LAMOST target image by AEM while the spectral traces are derived from
the corresponding flat field image. The size of each PSF is 15 pixels
$\times$ 15 pixels. Each input 1D spectrum (see Fig.\ref{inputspec})
is convolved with these PSFs along each spectral trace to generate a
2D spectrum for one fiber, then a 2D image with 250 fibers.
\par

In the computation process, the memory occupied is less than 300M Bytes, most
of which is used for storing the original and residual images. To avoid
singular matrix during computation when the viable block is too big,
we set the size of the viable block to 20.

\par
\begin{figure}

  \centering
   \includegraphics[width=126mm,height=178.2mm]{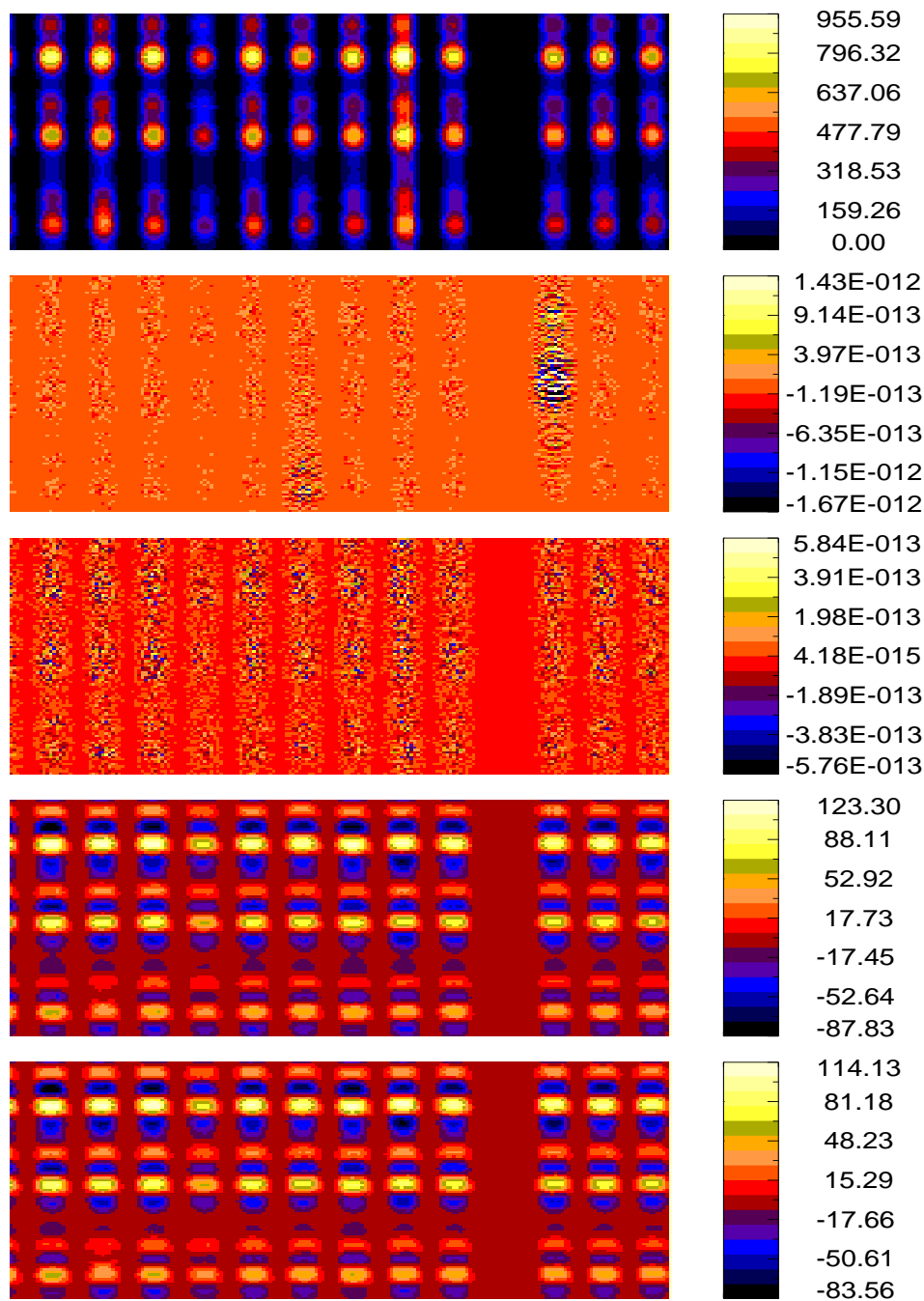}
  \caption{The top panel is a piece of simulation image, we apply different methods
   to this simulation image, the residuals are shown in  the lower 4 panels.
   From the 2nd to the bottom panels are the residuals of the DDA algorithm with $\epsilon = 10^{-2}$,
   the DDA with $\epsilon =10^{-4}$, the PFM and the AEM, respectively.
  Different colors represent count level on the CCD image as indicated by the color bars on the right.
  Compared with the PFM and the AEM, residuals of the deconvolution method
  are extremely small, especially around the region of emission lines.}

  \label{res}
\end{figure}
\begin{figure}

  \centering
   \includegraphics[width=126mm,height=178.2mm]{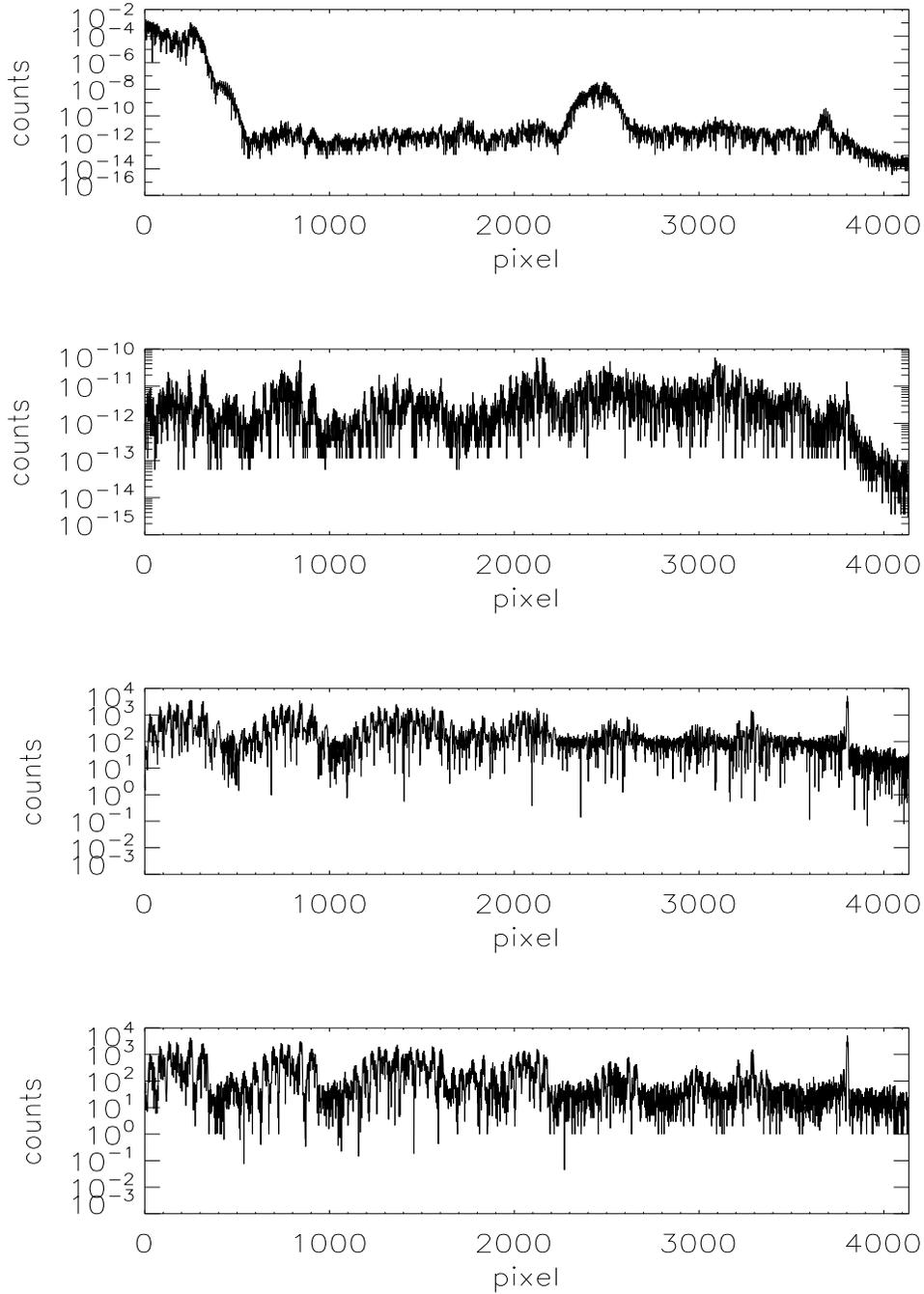}
  \caption{This figure shows the absolute values of the difference between the input spectrum and
  the spectrum extracted by different methods. From top to bottom, the methods are:
  the DDA with $\epsilon = 10^{-2}$, the DDA with $\epsilon = 10^{-4}$, the PFM and the AEM, respectively.
  Comparing with  the traditional PFM and AEM methods, we can see that the residuals of deconvolution
   method are negligible.
  \label{flux_diff}}

\end{figure}

Two tests are run following the steps described in section \ref{sec:Algor},
 the computation precision $\epsilon$s are set to $10^{-2}$ and
$10^{-4}$, respectively. After running the program for 7,074 and
17,989 seconds, the results can be seen in
Fig.\ref{res}. The top panel of Fig.\ref{res} shows the input
simulation image, the lower 2 panels show the residuals after
 the spectra are extracted with $\epsilon = 10^{-2}$ and $10^{-4}$
respectively. For comparison, we put the 2D residual of the PFM in the
fourth panel and the residual of the AEM in the bottom panel. From the
corresponding color bars we can see that the residuals are extraordinarily
small for our deconvolution method, and there are no obvious
residuals around the emission lines in the 2nd and the 3rd panels. In contrast,
the traditional PFM and AEM leave about 5$\sim$7\% residuals around
strong emission lines.  Fig.\ref{flux_diff} shows
 the absolute values of the differences between the input and the extracted spectrum.
The upper 2 panels show the residuals  of our
deconvolution method with different $\epsilon$. Only
one of the 250 spectera is shown to illustrate our results. We have
checked all the other spectra, they are all at the similar residual
level to the spectrum shown in Fig.\ref{flux_diff}. As we can see, the
residual is very small with about 2 hours
computation when we set $\epsilon$ to $10^{-2}$, and they are almost
negligible but with a longer computation time when $\epsilon$ is set to
$10^{-4}$. We also plot the residuals by the PFM and AEM in the
lower two panels. Compared with the much larger
residuals of the PFM and AEM, our method shows great advancement.
\par
As can be seen from the top panel of Fig.\ref{flux_diff}, the absolute
values of the residuals at the beginning part of the spectrum are
relatively larger than the residuals after the first 500 pixels.
This is because that the calculation begins from one end of each
spectrum in order of pixels. As described in \ref{sec:Algor},
in each process of calculation, there is a block of $\lceil k/2 \rceil$
pixels overlapping with the previous block. Results of these
$\lceil k/2 \rceil$ pixels in the previous block will be the
input of the corresponding pixels in the current block. This
guarantees that the solutions of conjunct blocks are related.
Thus residuals in the later pixels will be smaller since their
initial input had been processed closer to the final solution
than the beginning part. The residual of the beginning part
will be reduced if the program iterates long enough time,
as shown in the second panel of Fig.\ref{flux_diff}.

\subsection{Applied to simulations with cross talk}
\label{sim_crtk}

\begin{figure}

  \centering
   \includegraphics[width=126mm,height=178.2mm]{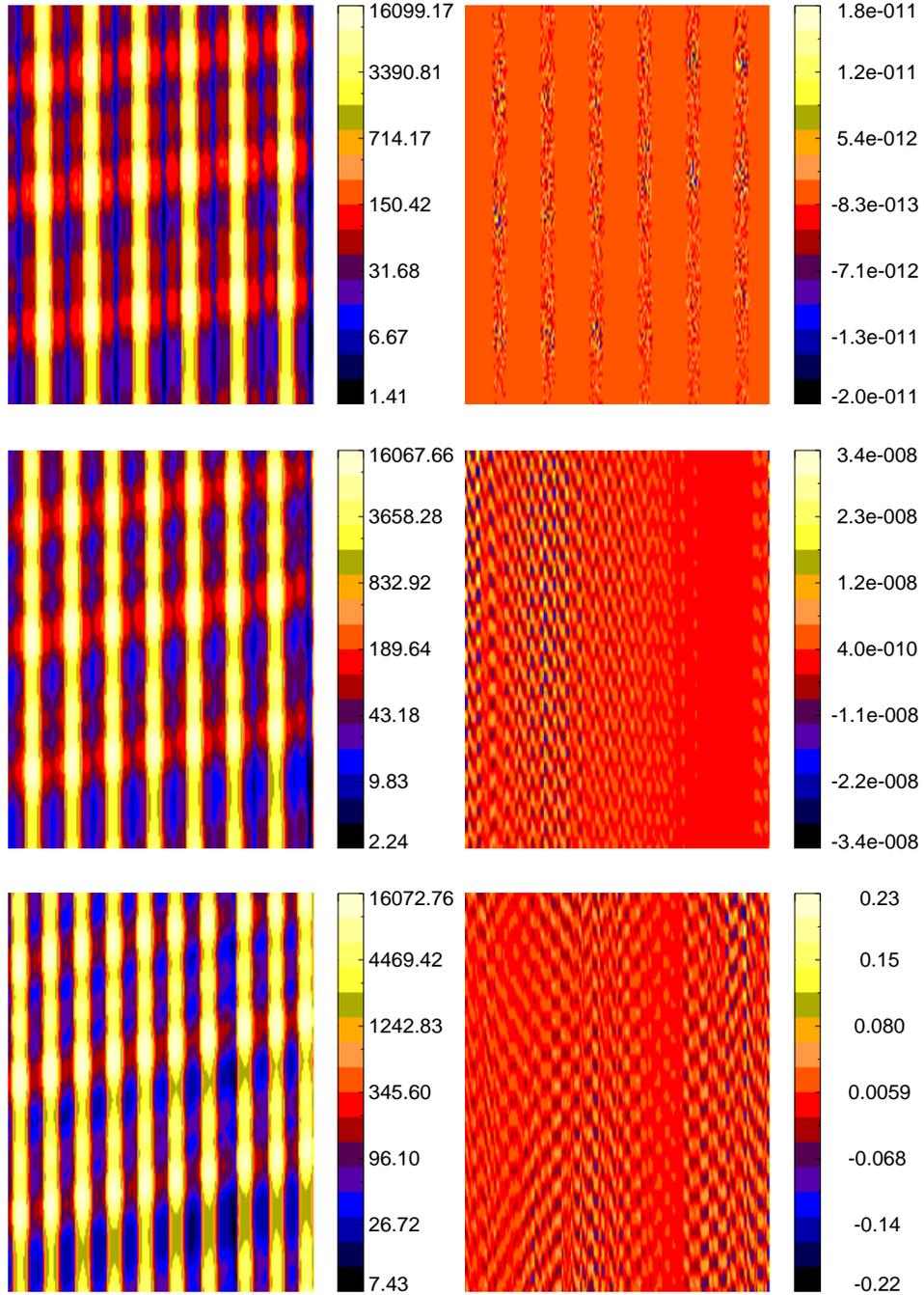}
  \caption{From top to bottom, the fiber distances in each row are set
  to 10, 8 and 6 pixels, respectively. Panels on the left column
  are  pieces of  simulated 2D images. Every 3 spectra, there is a
  spectrum 100 times stronger than its neighbors,  only the bright
  fibers can be seen due to high contrast. Panels on the right column
   are the  corresponding residuals after the algorithm runs for about 2 hours.
  Counts are represented by colors as indicated by color bars.
  \label{crosstalk-res}}

\end{figure}

In the above simulations, the distances between fibers were set to
the same as LAMOST, about 15-16 pixels. The size of each PSF is 15
pixels $\times$ 15 pixels and the FWHM (Full Width at Half Maximum)
is about 6 pixels, so fiber-to-fiber cross talk  in the above simulations
is very small. To test our algorithm on images with serious cross talk,
we construct 3 images using the same method  discussed above but with the distances
between fibers  shrunk to about 10, 8 and 6 pixels,
respectively (see Fig.\ref{crosstalk-res}). We put 250 fibers in each image, the size in
wavelength direction is kept as 4K pixels, so the image sizes are not the same for different fiber distances.
Since the FWHM of LAMOST PSF is about 6 pixels, we do not try to simulate distances less than that. To
study the influence of bright fibers on their neighbors,
every  3 spectra, there is a spectrum with its flux 100 times stronger than its neighbors.
\par
\begin{figure}

\vspace{2mm}
   \hspace{3mm}
   \includegraphics[width=128mm,height=178.2mm]{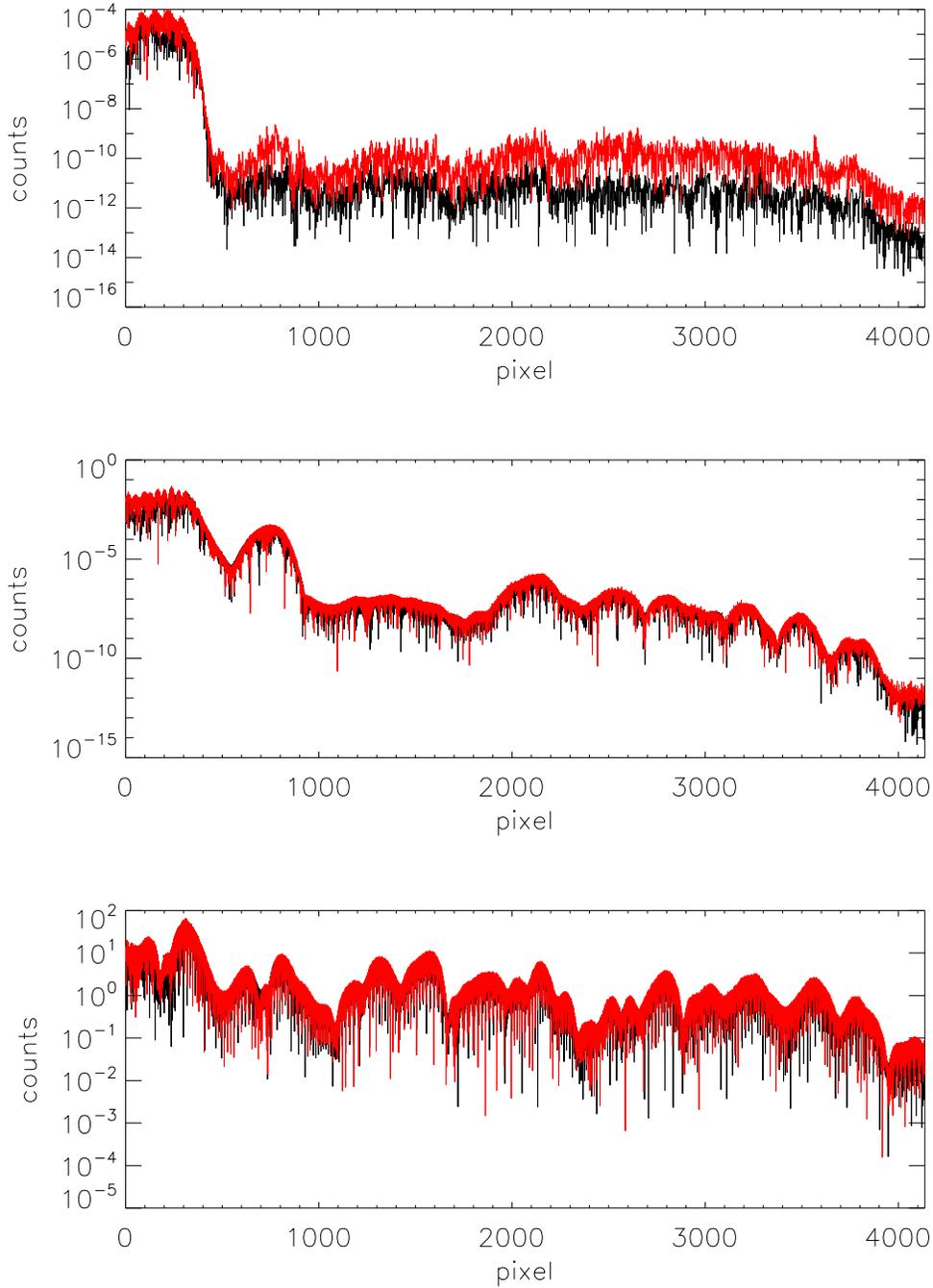}

    \caption{Absolute values of 1D residuals after the algorithm runs
    for about 2 hours ($\epsilon \sim 10^{-2}$). From   top to  bottom, panels illustrate
    absolute values of residuals of two 1D spectra extracted from 2D
    images with fiber distance about 10, 8 and 6 pixels, respectively.
    In each panel, the red spectrum is the residual
    of the bright fiber and the black spectrum is its  neighbor with flux 100 times weaker.
    \label{crosstalk-dif}}

\end{figure}
\begin{figure}

\vspace{2mm}
   \hspace{3mm}
   \includegraphics[width=126mm,height=178.2mm]{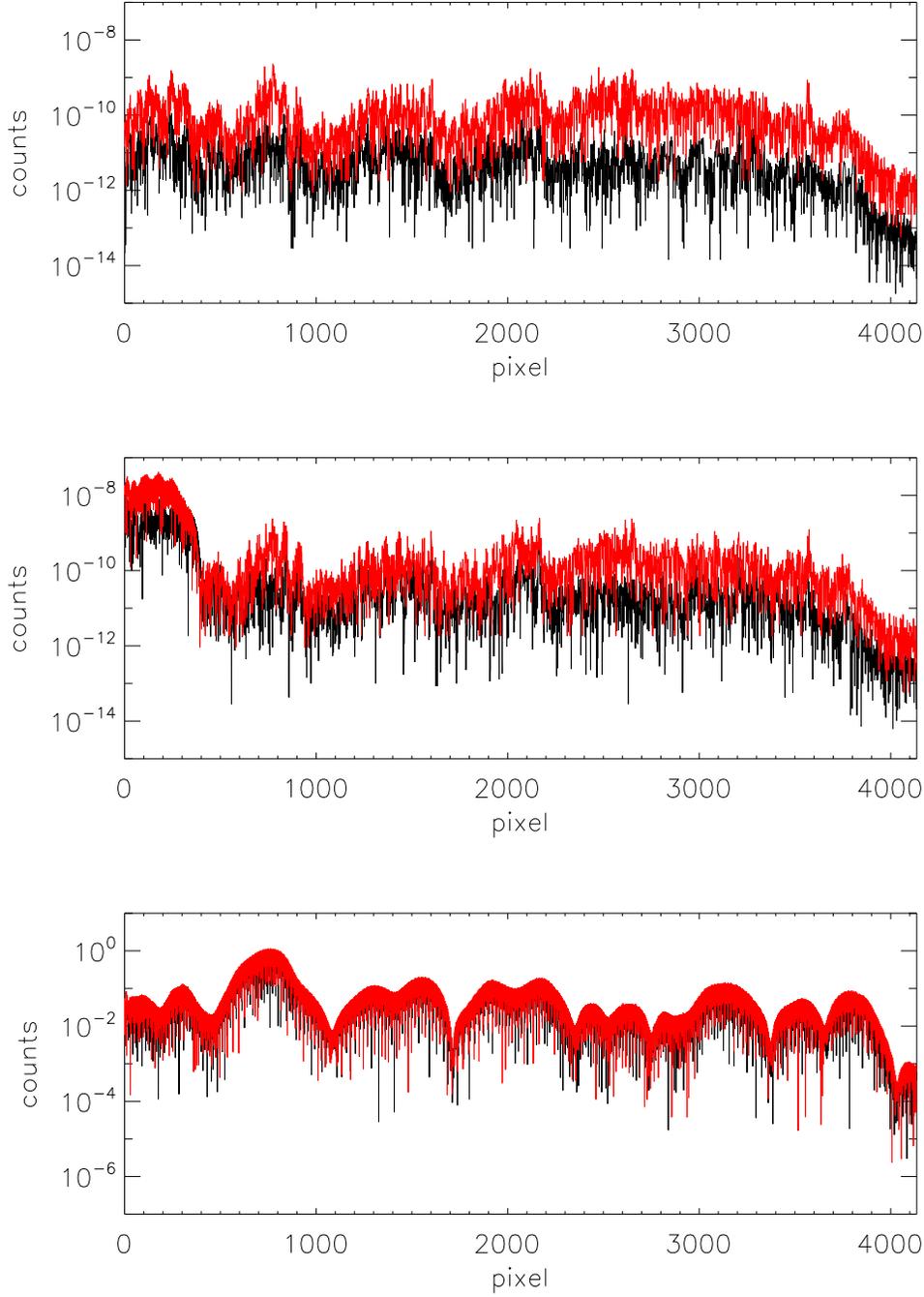}

    \caption{The same residuals as Fig.\ref{crosstalk-dif}, but for the
    program running for about 5 hours ($\epsilon \sim 10^{-4}$).
    \label{crosstalk-dif7}}

\end{figure}

 The program runs about 2 hours by setting $\epsilon$ to about $10^{-2}$, 2D and 1D residuals are shown
in Fig.\ref{crosstalk-res} and Fig.\ref{crosstalk-dif} respectively.
Only two of the spectra (a bright fiber and its faint neighbor) are shown in Fig.\ref{crosstalk-dif} to
represent the common situation. From these two figures we can see:
Firstly, both 1D and 2D residuals are small and acceptable. The residuals
increase when fiber distance decreases, but not in a linear tendency.
In fact, the residual grows much faster as the distance decreases to
the extreme case, 6 pixels. Secondly, residuals of the bright fiber and its
faint neighbor are on the similar level and the influence of the
bright fiber on its neighbor is negligible in our method.
Thirdly, as discussed in Section \ref{sim_data}, residuals at
the beginning parts of the 1D spectra (Fig.\ref{crosstalk-dif}) are
larger than in the rest parts. They could be reduced by setting
 computation precision $\epsilon$ to a smaller value but with longer execution time.
 As shown in Fig.\ref{crosstalk-dif7},
  residuals at the beginning part are greatly reduced after the
program runs about 5 hours when $\epsilon$ is set to about $10^{-4}$.

\subsection{Comparison with Bolton \& Schlegel's method}

\begin{figure}
	\includegraphics[height=180mm]{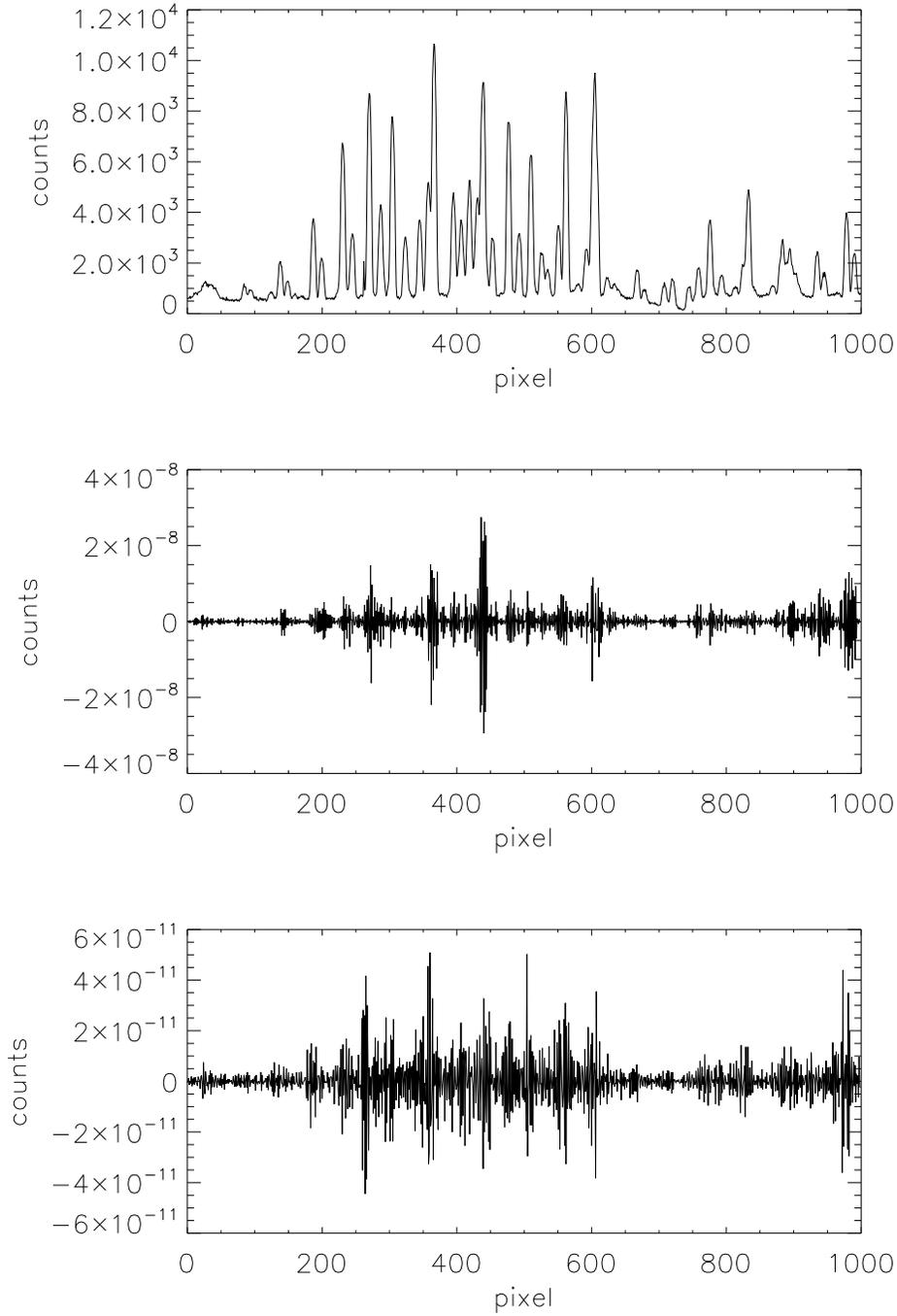}
    \caption{The top panel is the input 1D spectrum used to generate 2D image;
    the middle panel is the difference between the input spectrum and
    the spectrum extracted by Bolton \& Schlegel's method; the bottom
    panel is the difference between the input spectrum and the spectrum
    extracted by the DDA. Comparing the lower two panels, the accuracy of
    our result is much higher than Bolton \& Schlegel's.
    \label{bolton-cmp}}

\end{figure}

To compare with the method of Bolton \& Schlegel, a piece of 1D spectrum,
with only 1,000 pixels, is convolved with LAMOST PSFs to generate a 2D
image of one fiber. Then we use the DDA  and the method in \cite{bol10}  to
deconvolute the 2D image respectively. Results are shown in Fig.\ref{bolton-cmp},
 we can see that our method can achieve higher accuracy than that of \cite{bol10}.
When calculating an inverse matrix, calculation error is accumulated, so a larger matrix
leads to a bigger calculation error. The calibration matrix (see equation \ref{Guassian_solution}
and  \ref{der_proc}) in our method is much smaller than Bolton \& Schlegel's,
so the calculation accuracy of its inverse can be higher  than theirs. Meanwhile, calculation of large
calibration matrix is very time consuming. It tooks about 123 seconds for  Bolton \& Schlegel's method to run, while only 3 seconds for the DDA. Therefore,
the DDA has better performance in both calculation accuracy and computation time.

\subsection{Applied to  simulations  with poisson noise\label{sec:44}}

In the above simulations, noise is not considered, yet in actual observations, noise is
inevitable. Deconvolution is an ill-posed problem,  its result is very
sensitive to noise.  From the upper 2 panels of Fig.\ref{poisson_1d_cmp} we can
see that if we use the DDA to directly extract an image with noise,
the spectrum will be dominated by noise. Many people (e.g. \cite{sta02},
\cite{pue05}, \cite{dav12} and references therein)
have discussed how to minimize the noise influence. Here we will
use \cite{tik63} and \cite{tik77} regularization methods to control the
noise. Introducing the first derivative as regularization item
in the Tikhonov method, function \ref{obj} could be rewritten as:
\begin{equation}\label{obj_reg_dir}
 \min_{c_{i,j}} \sum^{N}_{k=1} \sum^{M}_{m=1}(F[k,m] -
\sum^{N_f}_{i=1} \sum^{N}_{j=1} c_{i,j}PSF_{i,j}[k,m])^2 + \sum^{N_f}_{i=1} \sum^{N-1}_{j=1} \alpha_{i,j}(c_{i,j+1}-c_{i,j})^2,
\end{equation}
where $\alpha_{i,j} \geq 0$ is a weight to adjust the regularization item $(c_{i,j+1}-c_{i,j})^2$.
We can rewrite equation \ref{obj_reg_dir} in matrix form:

\begin{equation}\label{obj_reg_mat}
 \min_{C} (F-AC)^T(F-AC) + (\Gamma C)^T \alpha(\Gamma C),
\end{equation}
where $\Gamma$ is the Tikhonov matrix consisting of 250 submatrix blocks on
its diagonal. In each submatrix, the diagonal element $(k,k),
(k=1,2, ..., N-1)$ is -1, the element $(k,k+1),(k=1,2, ..., N-1)$ is 1,
and all the other elements are 0. $\alpha$ is a diagonal matrix consisting of
250 submatrix blocks, in the $i$th submatrix block, the diagonal element
$(j,j),(j=1,2, ..., N-1)$ is $\alpha_{i,j}$.
\par

 For an initial guess $C_0$, we define the same variables: $\delta C$,  $X_s$, $\bar{A}$,
 $R$ and $res$ as in section \ref{sec:Algor},
 define the $s+1$ th, $..., s+k$ th columns in $\Gamma$ as submatrix $\bar{B}$,
then $A\delta C \equiv \bar{A}X_{s}$, $\Gamma\delta C \equiv \bar{B}X_{s}$.
Substituting the variables  above into objective function \ref{obj_reg_mat}, we could write the function in a new form:
\begin{flalign}
\label{tder_proc_Tikh}
\begin{split}
            & (F-A(C_0 + \delta C))^T(F-A(C_0+\delta C)) + (\Gamma (C_0+\delta C))^T \alpha(\Gamma (C_0+\delta C)) \\
             = &  (R - \bar{A}X_{s})^T(R -\bar{A}X_{s}) + (\Gamma C_0+\bar{B}X_{s})^T \alpha(\Gamma C_0+\bar{B}X_{s})\\
             =  & res - 2R^T\bar{A}X_{s}+ X_{s}^T\bar{A}^T\bar{A}X_{s} + (\Gamma C_0)^T \alpha(\Gamma C_0) + 2 C_0^T\Gamma^T\alpha\bar{B} X_{s} + X_{s}^T\bar{B}^T\alpha\bar{B}X_{s}\\
           = & res + (\Gamma C_0)^T \alpha(\Gamma C_0) - 2(R^T\bar{A} - C_0^T\Gamma^T\alpha\bar{B}) X_{s}  + (X_{s}^T\bar{A}^T\bar{A}X_{s} + X_{s}^T\bar{B}^T\alpha\bar{B}X_{s}).
%
\end{split}
\end{flalign}
If we  let $X_s = (\bar{A}^T\bar{A} + \bar{B}^T\alpha\bar{B})^{-1}(\bar{A}^TR - \bar{B}^T \alpha\Gamma C_0)$, where the objective function
achieves its local minimum in the block $X_s$, and denote $ RES = res + (\Gamma C_0)^T \alpha(\Gamma C_0)$, then the value of the objective
function becomes:
\begin{equation}\label{func:RES}
RES - (\bar{A}^TR - \bar{B}^T \alpha\Gamma C_0)^T (\bar{A}^T\bar{A} + \bar{B}^T\alpha\bar{B})^{-1}(\bar{A}^TR - \bar{B}^T \alpha\Gamma C_0) \leq RES.
\end{equation}
Similar to section \ref{sec:Algor}, the second term in the left side of the function is a non-negative quadratic term,  so the objective function will keep decreasing for
every small block $X_s$ in each
iteration until the decrease is smaller than a given precision, the objective function will finally converge close to its global minimum.
Because $\Gamma$ is a sparse matrix with non-zero elements all on the primary
and the higher second diagonal, and $\alpha$ is also  a sparse matrix with
non-zero elements all on the diagonal, thus $\bar{B}^T\alpha\bar{B}$ and $\bar{B}^T \alpha\Gamma C_0$ can be sparsely stored and calculated.

During the computation process, $\Gamma_d$ denotes the decrement of
the Tikhonov item in each iteration, $r_1$ denotes the decrement of $res$, other symbols are the
same as defined above. We then
change the DDA slightly, and rewrite it as following:
\par

Step 0: Given a value $\epsilon > 0$, the size of block: $k$, and
the size of variable $C$: $N$, $0 < k \ll N$;
\par
Step 1: Let $C_0 := 0$, $s := 0$. For function \ref{obj_reg_mat},
let $R := F$, $res := R^TR$, $\Gamma_d := 0$, $r_1 :=0$, $RES := res$;
\par
Step 2: If $s + k > N$, $s := N$, goto setp 3; else let $\bar{A}$
be the submatrix consisting of the $(s+1)$th$, ...,(s+k)$th columns
in $A$, and $\bar{B}$ be a submatrix consisting of the $(s+1)$th,
$..., (s+k)$th columns in $\Gamma$. We denote $X_s =
(\bar{A}^T \bar{A} + \bar{B}^T\alpha\bar{B})^{-1}(\bar{A}^TR -
\bar{B}^T\alpha \Gamma C_0)$. Let $r_1 := r_1 + 2R^T\bar{A}X_{s} -
X_{s}^T\bar{A}^T\bar{A}X_{s}$ ; $\Gamma_d:= \Gamma_d
-(2C^T_0\Gamma^T\alpha\bar{B}X_s + X^T_s\bar{B}^T\alpha\bar{B}X_s)$;
$(c_{s + 1},...,c_{s + k})^T := (c_{s + 1},...,c_{s + k})^T + X_s$;
$R := R - \bar{A}X_{s}$ and $s := s + \lceil k/2 \rceil + 1$, goto step 2;
\par
Step 3: If $s - k < 0$, $s := 0$, then go to step 4; else let
$\bar{A}$ be the submatrix consisting of the $(s - k + 1)$th
$,..., s$th columns in $A$,  $\bar{B}$ be a submatrix consisting
of the $(s - k + 1)$th $,..., s$th columns in $\Gamma$. We denote
$X_s = (\bar{A}^T\bar{A} + \bar{B}^T\alpha \bar{B})^{-1}(\bar{A}^TR
- \bar{B}^T \alpha\Gamma C_0)$. Let $r_1 := r_1 + 2R^T\bar{A}X_{s} - X_{s}^T\bar{A}^T\bar{A}X_{s}$;
 $\Gamma_d:= \Gamma_d -(2C^T_0\Gamma^T\alpha\bar{B}X_s + X^T_s\bar{B}^T\alpha\bar{B}X_s)$; $(c_{s - k + 1},...,c_{s})^T :=
(c_{s - k + 1},...,c_{s})^T + X_s$; $R := R - \bar{A}X_{s}$ and
$s := s - \lceil k/2 \rceil - 1$;

\par

Step 4: $res := res - r_1$ and $RES := RES - r_1 - \Gamma_d$; If $r_1 + \Gamma_d< \epsilon$, stop; else let $r_1 = \Gamma_d = 0$,  goto step 2.
\par

We call this algorithm  \emph{the Tikhonov Deconvolution Algorithm}, or TDA, here after.
\par

\begin{figure}

   \plotone{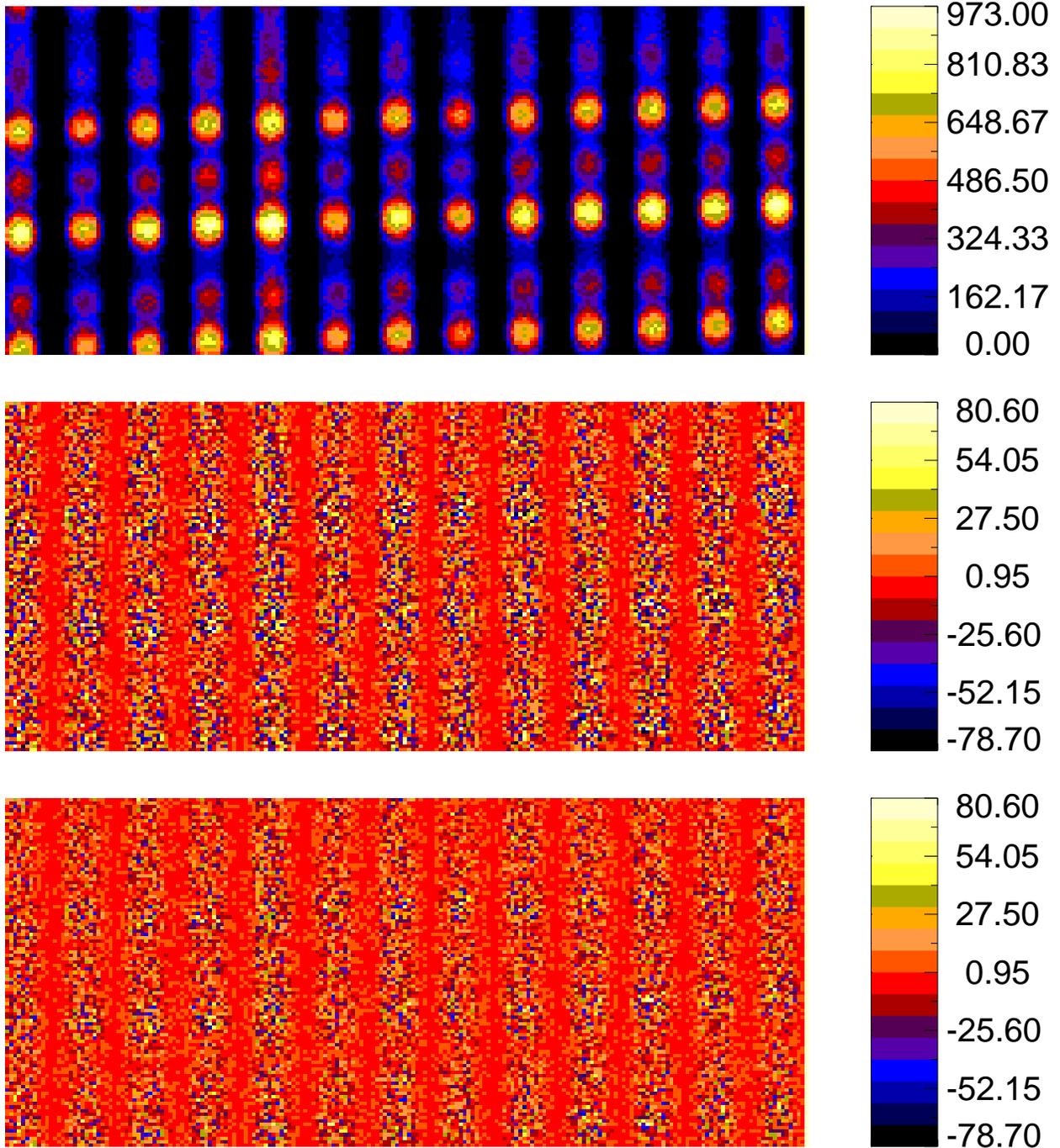}
    \caption{The top panel is the simulation 2D image with poisson noise,
    the middle panel is the residual of the DDA and the bottom panel
    is the residual of the TDA. Pixel counts are represented by different
    colors, as indicated by color bars on the right. The mean  of
    the DDA and the TDA residuals are  -0.0028 and 0.0026 respectively, and the standard deviations of the DDA and the TDA residuals are 12.23 and 12.75
    respctively, so the 2D residuals of the two methods are almost the same.
    \label{rescmp-poisson}}

\end{figure}

\begin{figure}[p]

   \includegraphics[width=12cm]{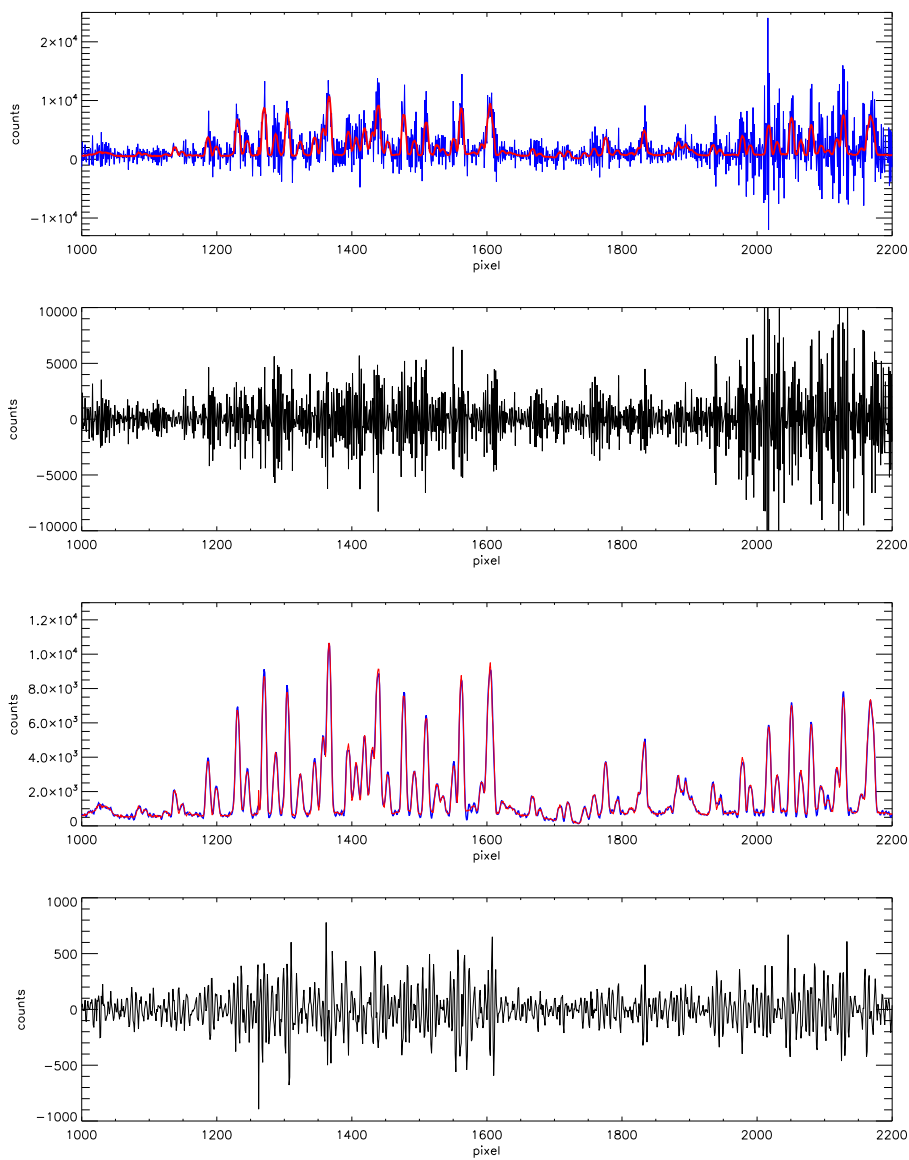}
    \caption{From  top to  bottom: the first panel shows the input spectrum (red curve) and
    the spectrum extracted by  the DDA from the noise-added image shown in Fig.\ref{rescmp-poisson} (blue curve); the second panel shows the difference between
    the two spectra in the first panel, from which we can see the
    influence of noise on the DDA is enormous; the third panel shows the  spectrum extracted by the TDA (blue curve) and the same input spectrum as in the top panel (red curve);
    the last panel shows the difference between the two spectra in the third panel. Comparing the second panel
    with the fourth panel, noise introduced by the DDA is greatly suppressed
    by the Tikhonov regularization. To make a close view of the spectra, only a part of the full wavelength coverage is displayed here.
    \label{poisson_1d_cmp}}

\end{figure}

To test the TDA, we generate  a simulation image similar to that in Section
\ref{sim_data}, except that poisson noise is added to each pixel on the 2D
 image. Then the 1D spectra are extracted  by the DDA and the TDA.
 In the TDA,  all the diagonal elements of $\alpha$ are set to 0.001. Results are shown in Fig.\ref{rescmp-poisson} and  Fig.\ref{poisson_1d_cmp}. 2D residual images
 in Fig.\ref{rescmp-poisson} show little difference between the DDA and the TDA,
 while compared with the DDA, the noise level in the 1D TDA result  is greatly suppressed
 by the Tikhonov regularization, as show in Fig.\ref{poisson_1d_cmp}.
\par

\par
\begin{figure}

   \plotone{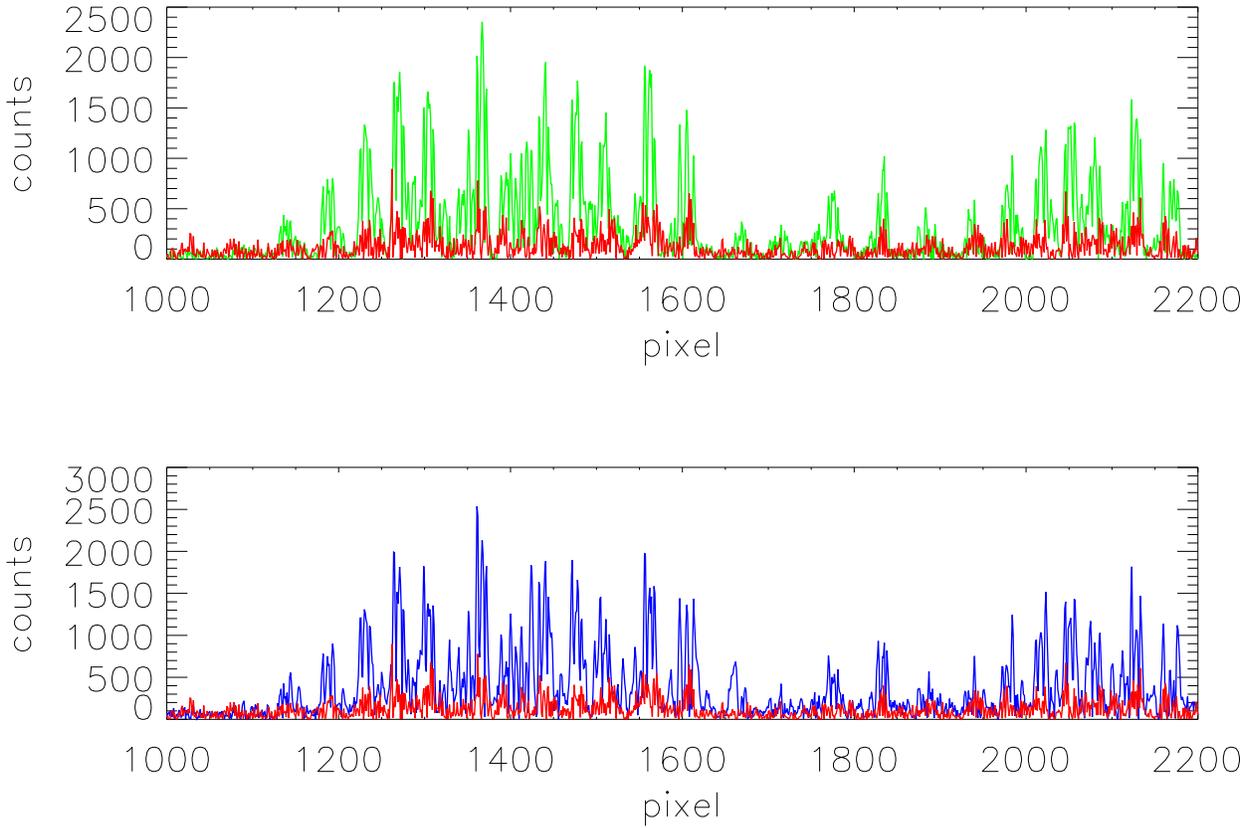}
    \caption{To compare the residual of the spectra extracted from the noise-added image by different extraction
    methods, we show the absolute value of the AEM residual (green curve) in the upper panel
     and the absolute value of the PFM residual in the lower panel (blue
    cure), the absolute value of the TDA residual are plotted in red in both panels.
    Compared with the AEM and the PFM, the residual of the TDA is the smallest.
     \label{res_cmp}}

\end{figure}

We also extracted the noise-added 2D image  by the AEM and the PFM.
For comparison,  absolute values of 1D residuals of different methods are shown in
Fig.\ref{res_cmp}. From the figure, we can
see, compared with the AEM and the PFM, residual of the TDA is  smaller, so
1D spectra extracted by the TDA are the most reliable.

\begin{figure}

   \plotone{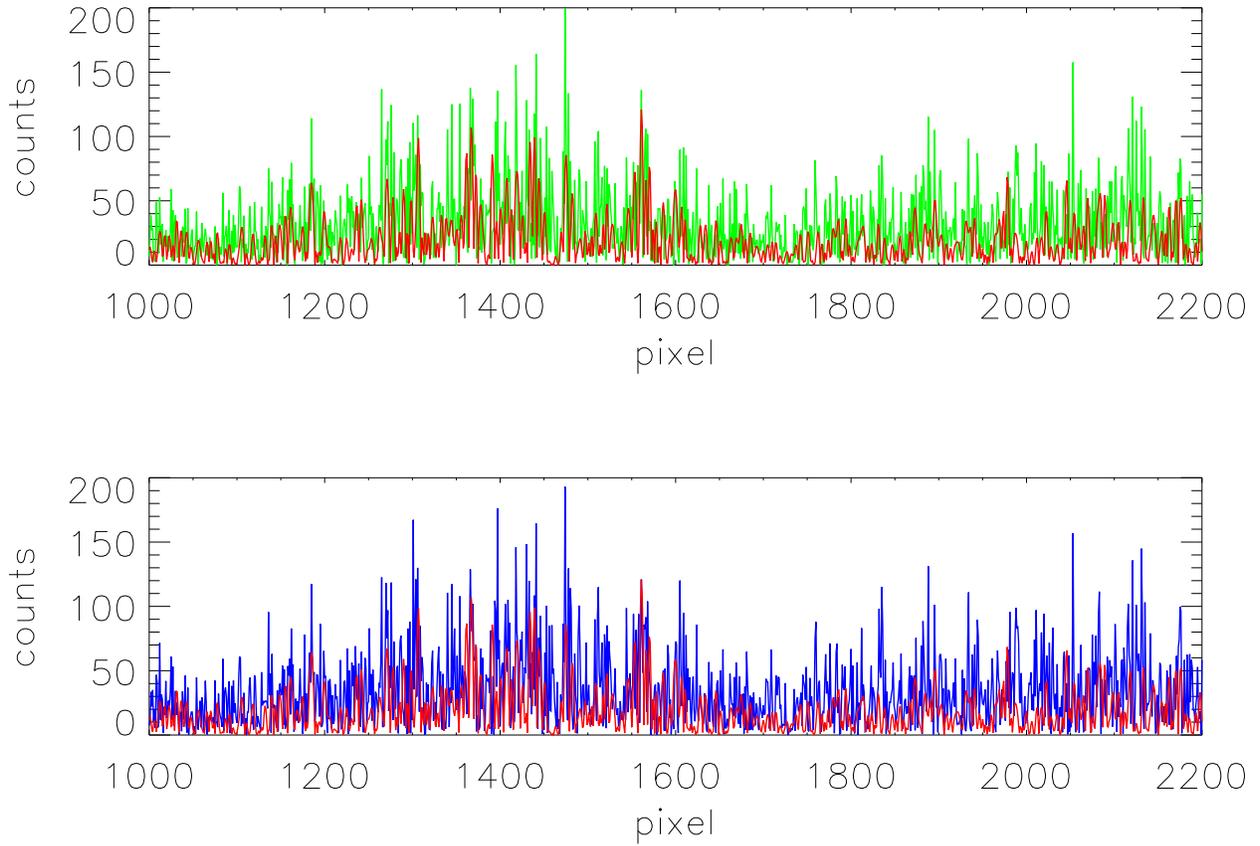}
    \caption{The same residual plots as Fig.\ref{res_cmp}, but to compare the noise level of different extraction methods
    in the same resolution, both the input spectrum and the spectrum
    extracted by the TDA are degenerated to the resolution of the AEM and the PFM by convolving
    with a gaussian profile.
    Compared with the AEM and the PFM, the influenced of noise in the  TDA is much less even if the resolution is the same.
    \label{noise_cmp}}

\end{figure}

 In the simulation image, the distances between LAMOST fibers(15-16 pixels)
 are the same as the size of the input PSF (15 pixels), then  uncertainty
 caused by the fiber cross talk is quite small, so the noise in the TDA 1D spectra
 should mostly come from the method in extracting the 2D image with poisson noise.
 But since the TDA could extract the spectrum with the similar resolution to the input spectrum, whereas the traditional methods could not,
 so the large residuals in the AEM and PFM are partly from the resolution difference between the input and the output spectra.
 To compare the noise level of different extraction methods under the
 same resolution, both the input spectrum and the spectrum extracted
 by the TDA are degenerated to the resolution of the AEM by convolving with a gaussian
 profile. Then the degenerated input spectra are subtracted from the 1D
 spectra  extracted by different methods.
 In Fig.\ref{noise_cmp}, the absolute values of residuals of the AEM, the PFM and the TDA are plotted in green,
 blue and red,  respectively. From the
 figure we can see, the residual of the TDA is obviously smaller than those
 of the AEM and the PFM. Therefore, our Tikhonov regularization item could successfully reduce the noise influence.
\par

\subsection{Applying the TDA to  simulation images with poisson noise and cross talk\label{NoiC}}
\begin{figure}

   \includegraphics[height=14cm]{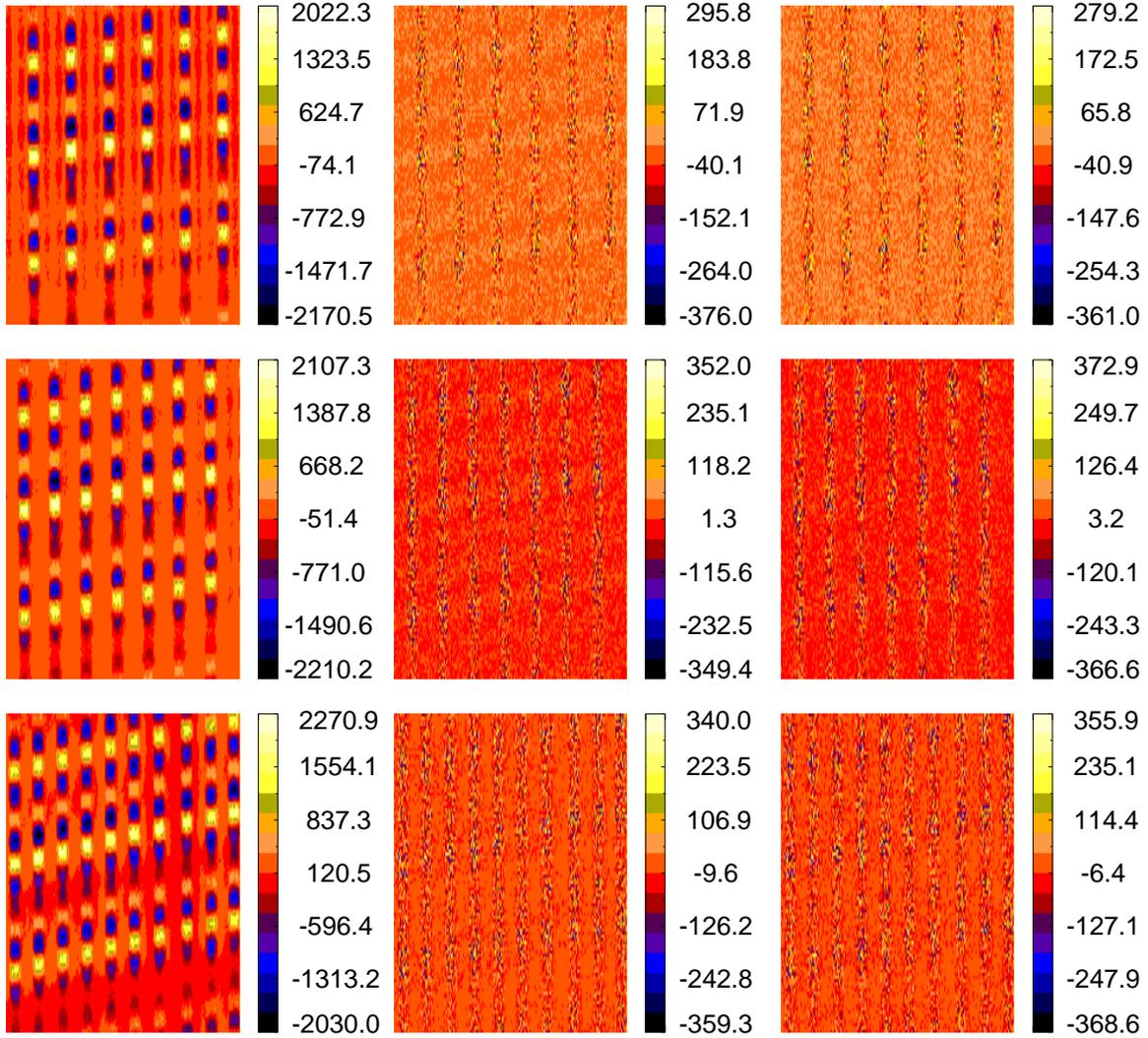}
    \caption{From left to right, columns are 2D residuals of the PFM,
    2D residuals of the TDA and the input 2D poisson noise,respectively. From  top  to
    bottom,  the distances  between fibers in different rows are 10, 8 and 6 pixels,
    respectively. Comparing the medium to the right column, the  2D TDA residuals are
    at the similar level of the poisson noise.
    \label{2D_poisson_sim_res}}

\end{figure}

\begin{figure}

   \includegraphics[height=16cm]{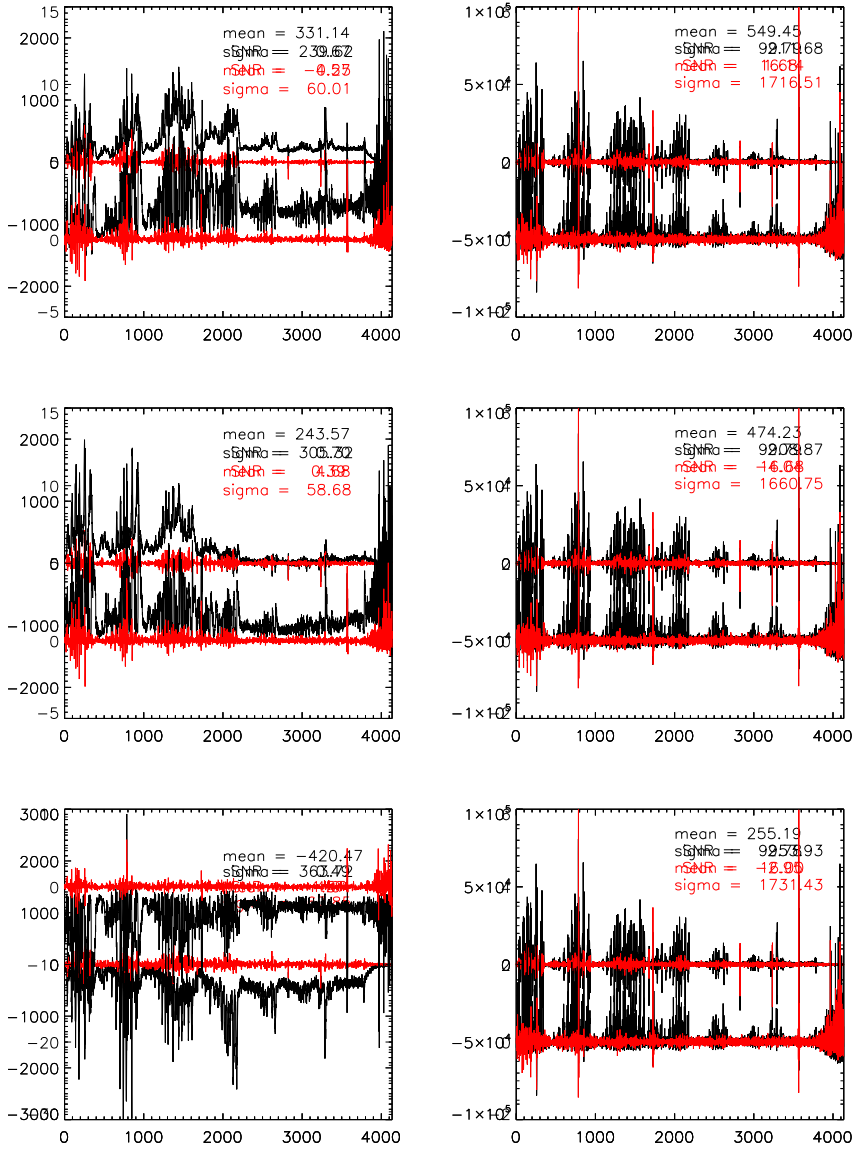}
    \caption{ 1D residuals from images in Fig.\ref{2D_poisson_sim_res}. The left and right columns show
    1D residuals of one of the bright fibers and its faint neighbor respectively. From  top
    to bottom, the distances between fibers are
    10, 8 and 6 pixels, respectively. In each panel, the red spectrum is the residual of the TDA
    and the black is the residual of the PFM. We also calculate the average SNRs of the extracted spectra,
    which are marked with the corresponding colors in each panel.
    \label{1D_poisson_sim_res}}

\end{figure}

To test our method under more stringent conditions, we  add poisson noise to the 2D images used in subsection \ref{sim_crtk}, on which the
 fibers  are separated by about 6, 8 and 10 pixels,
respectively. We extract the noise-added images by the TDA. The weights of
the Tikhonov item, $\alpha$,  are set to 0.0003 and 0.03 for the bright and the faint fibers
respectively, the calculation precision $\epsilon$ is set to $10^6$.
 The 1D and the 2D residuals are shown in
Fig.\ref{2D_poisson_sim_res} and Fig.\ref{1D_poisson_sim_res}
respectively. For comparison, the residuals of the PFM are also plotted in these figures.
We also list the corresponding SNR of each spectrum in Fig.\ref{1D_poisson_sim_res}, here, the SNR is defined as the mean of the extracted spectrum divided by the
mean square root of the difference between the extracted spectrum and the simulation input spectrum.
 From these figures we can see: Firstly, the 2D residuals of the TDA are comparable to poisson noise, which means that our
can preserve exactly all fluxes recorded by the CCD.
Secondly, both the 1D and 2D residuals of the PFM are larger than those of the TDA, the TDA spectra show much higher SNR than the PFM.
 Thirdly, as shown in Fig.\ref{1D_poisson_sim_res}, the 1D spectra of faint fibers are severely polluted by their bright neighbors  in all the PFM results,
 meanwhile, the corresponding results of the TDA show that the  residuals of both faint and bright fibers are around zero.
 The residuals in our TDA method show no strong increase with the decreasing distance between fibers.
 From these simulation results we can see,   the TDA is much more reliable than the traditional methods especially when there is cross talk.

\subsection{Computation time\label{sec:cputime}}
  The computation times in the above simulations are 167m, 168m and 252m for fiber distances of 10, 8 and 6 pixels, respectively.
  As may be noticed, the calculation precision $\epsilon$ is set to $10^6$ in section \ref{NoiC}, which is much larger than $10^{-2}$ in the noiseless simulation
  in section \ref{sim_crtk}. As defined in section \ref{sec:Algor} and \ref{sec:44}, computation precision $\epsilon$ is the threshold of the decrease
   of  the objective function,
  which is summed over all pixels, that is, for our simulation with 250 fibers and 4k pixels in wavelength direction, the total number of pixels is much more than $10^6$ (the accurate number depends on the fiber distance, or how many pixels the 250 fibers occupy), so $\epsilon=10^6$ means we have an average difference much less
    than 1 per pixel .
    For the noise free simulation in section \ref{sim_crtk}, our purpose is to test if our DDA method could work correctly with fiber-to-fiber cross talk, $\epsilon$ is set to a small value to see if the result precision could be  arbitrarily high when the calculation time is long enough. For the noise  added simulation in section \ref{NoiC},
   the calculation precision is noise dominated, so  $\epsilon$  should be comparable to the noise  level rather than the unnecessarily small
   value in section \ref{sim_crtk}.  We have made another test with  $\epsilon$ set to $1.0e-20$, the results show that both the 2D and 1D residuals are at a similar level to that of $\epsilon=10^6$, but the computation time is 4 times longer.
  Beside  hardware,  computation time  should depend on the image quality (e.g. noise, fiber number and fiber distance, etc) and input calculation parameters such as $\epsilon$, usually it is  a case by case problem. It is better that we could know how to adjust the parameters to  balance  the computation time against the  result precision. To better understand
  how the result precision changes with different conditions,
   we set $\epsilon$ to an arbitrarily low value, i.e., $1.0e-20$, then for each iteration, we output the computation time, the change of RES(see section \ref{sec:44}), the SNRs of bright and faint fibers
  for the noise-added images of fiber distance 15, 10, 8  and 6 pixels respectively in Table \ref{tab:cputime}.  Only the first 9 iterations are shown in the table since the SNRs stop to increase for more iteration.  From the table we can see, the computation time of each iteration  are almost the same for different fiber distances, this is easy to understand since all the images we simulated contain the same number of fibers and the same size(4096 pixels) in wavelength direction,  the amount of calculation is proportional
   to the product of the fiber number and the size in the wavelength direction, so the computation times are almost the same. For  fiber distances of 15, 10 and 8 pixels, the result precision of larger fiber distance converges faster than the smaller ones, after the precision reach photon noise level ($\sim10^6$), they all converge at the similar speed.   But the precision
    decreases much slower for the extreme case, 6 pixels. If we use $\epsilon=10^6$ as the threshold for our computation, the 15, 10 and 8-pixels converge at a similar time, while the
    6-pixels will take about 50\% more time to reach the threshold, we highlight the computation times and the precisions in the table where they reach $10^6$ threshold. In the right part of the table, we list the SNRs of one bright fiber
    and its faint neighbor, the SNR is the same as defined in section \ref{NoiC}. We have checked the SNRs of other fibers, the SNRs of faint fibers in the first 2 iterations change from fiber to fiber, which reflect the fact that the first result depends on the initial guess and is not
     stable, then the SNRs of different fiber distances increase quickly to a similar level after the first 2 iterations. The SNRs for  other bright fibers are similar to those listed in the table.  From the table we can see that the SNR converges to a fix value even though the calculation precision keeps decreasing after each iteration, for the fiber distances of 15, 10, 8 and 6 pixels, the SNRs converge after 3, 3, 4 and 6 iterations, respectively. The last column in Table \ref{tab:cputime} shows the SNRs of faint fibers with distance 6 pixels, we can see that the SNR
     increases much slowly after the 4th iteration, the corresponding precision is 6.89e8,  comparing  the precision and the computation time to those of 15, 10 and 8 pixels in the table, we can see that  a better  $\epsilon$  to balance  computation time against precision may lie between $10^7$ and $10^8$.      From those tests we can see that for an image with noise, we could set $\epsilon$ to a value that the average precision in each pixel does not exceed  noise, usually 1 per pixel should be a lower limit.  Once the precision threshold is set, except for the extreme
     cross talk, our method converge after 2 to 3 iterations, so the computation time does not depend so much on fiber distance for most cases.

\subsection{Applying the TDA to real LAMOST data}
\begin{figure}

   \plotone{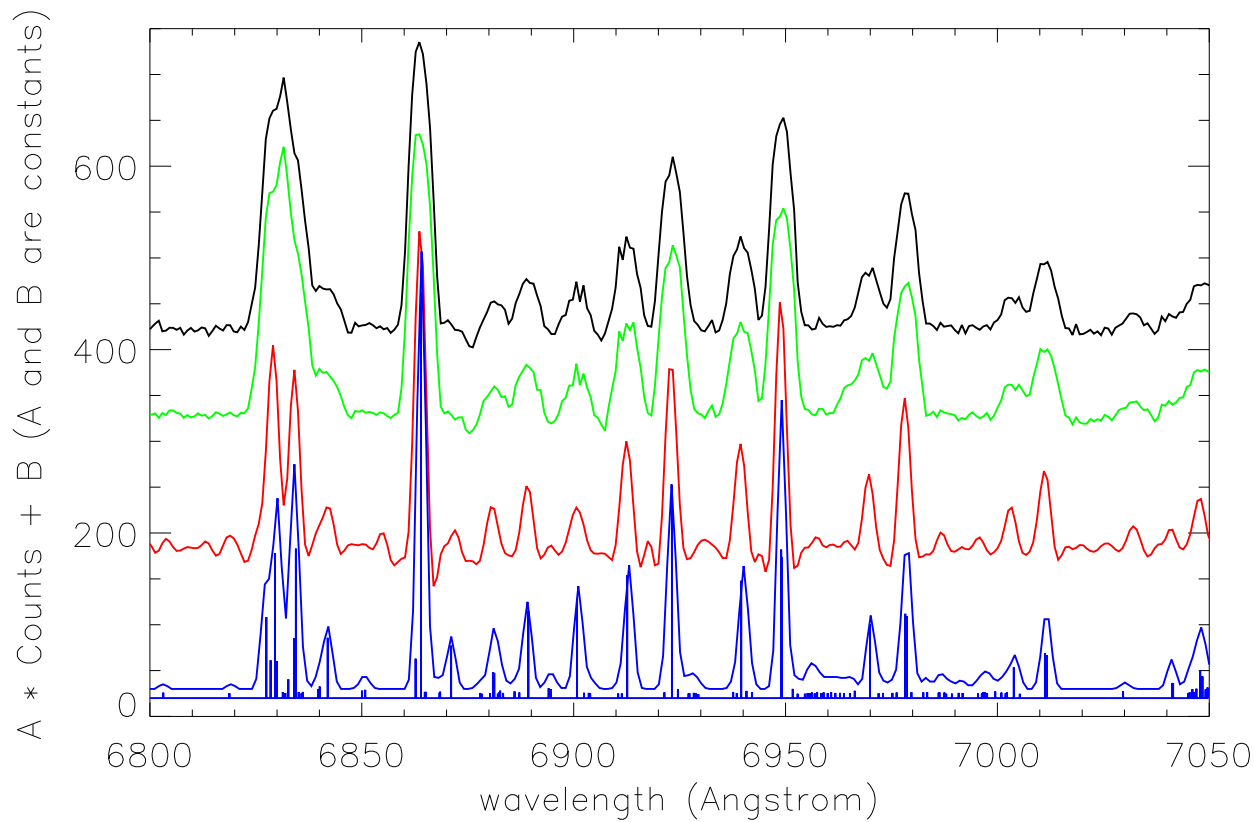}
    \caption{Results of extracting an actual LAMOST image.
    The spectra extracted by the AEM, the PFM and the TDA are plotted in black, green and red, respectively.
    The blue spectrum is a composite spectrum by convolving a gaussian
    profile with sky emission lines fluxes from \cite{skyemi}. We also plot
   the sky emission lines with  relative fluxes and positions from
    \cite{skyemi} in the bottom. Each spectrum is scaled and
    offset arbitrarily in y axis for clarity. 
    \label{reg_conv}}

\end{figure}

Now we apply our TDA to extract an actual LAMOST image.  Fig.\ref{reg_conv}
shows  the 1D spectra extracted  by the AEM(black curve),
the PFM(green curve) and the TDA(red curve)  from the same LAMOST
2D image, respectively. Most of the features in the spectra are sky emission lines.
 We also plot a composite spectrum (blue curve)
by convolving a gaussian profile ($\sigma$ =1 pixel, $\sim 0.7$\AA \
for LAMOST ) with the flux of sky emission lines from \cite{skyemi}. Relative fluxes and positions of those sky emission
lines are indicated by vertical blue lines in the bottom.

 From Fig.\ref{reg_conv} we can see:

        Firstly, some  emission
 lines  indiscriminate in the AEM and the PFM results are clearly resolved by our
 method, for example, the strong emission line in the black curve
 around $\lambda$6830  is clearly separated into two peaks in the
 red curve, which are OH $\lambda$6829.5 and a blend line of OH $\lambda$6832.7 and $\lambda$6834.4. Other lines like
 $\lambda$6842.0, $\lambda$6871.1,  $\lambda$6894,  $\lambda$7041.5
 and emission lines between  6975\AA \ and 7000\AA  \ are all
 clearly resolved by our method, as can be seen by comparing to  the blue composite
 spectrum. Compared to the composite spectrum, the TDA result has similar resolution(R$\sim$4200), which is 3 times higher than the AEM and the PFM results(R$\sim$1000).

\par

Secondly, some ambiguous bifurcate peaks in the AEM and the PFM such as
$\lambda$6912.6,  $\lambda$6900.8,  $\lambda$6969.9 and  $\lambda$7003.9
are explicitly detected as single peaks by our method, which indicates
that our result has higher SNR than those of the AEM and the PFM.

\par

Thirdly, carefully comparing the red spectrum with the black, green and blue
spectra, we find some weak suspicious emission  lines around strong
peaks, such as $\lambda$6855, $\lambda$6918 and $\lambda$6943, we think these features are most possibly caused by our  Tikhonov regularization method, although the features of the underlying object can not be completely ruled out.
Current Tikhonov item  we adopted may produce wavy features around strong peaks, the influence of noise could not be completely eliminated.
 We will leave how to further constrain the noise influence in the
deconvolution method for a future work.

\section{Conclusion and Discussion}
In this paper, we present a practical calculation scheme to extract
1D spectra from 2D images. Because of instrument optical aberration,
PSFs at different positions of a CCD image are mostly different. To sample PSFs,
we use emission lines on the arc image to generate
discrete basic PSFs, then we use B-spline surfaces to represent the smooth
basic PSF contours. By interpolating these basic PSFs,  PSFs at non-emission-line region can
be calculated. During the calculation process, only the smooth basic PSFs are stored in the memory,
other PSFs can be calculated when necessary, so the requirement of memory is reduced.
\par
Due to its huge requirement of computation resources,  it is
impossible to solve the objective function directly by the
least-square method described in \cite{bol10}. To solve this
problem, we try to reduce the objective
function in a small variable block in one calculation step, thus only a small amount of memory is needed.
The objective function could converge to its global minimum by gradually decreasing the variable blocks to
their local minimum. Thus, our
calculation scheme can solve the problem with common calculation resources.
\par

Based on our calculation scheme, we apply the Direct
Deconvolution Method to extract simulation 2D images without
noise. Results show that our methods could  extract the 1D spectra in a precision that traditional methods could not achieve,
even if fiber-to-fiber cross talk is significant.
\par
Deconvolution methods are sensitive to noise. To suppress the noise
influence, we introduce a Tikhonov item into the objective function.
Then we apply the TDA to simulations with poisson noise. Results show that the TDA can
extract the most reliable 1D spectra compared with the AEM and
the PFM. It can correctly extract the spectrum with extreme cross talk, even
for the faint fibers with fluxes 100 times lower than their neighbors.

\par

We also apply the TDA to an actual LAMOST 2D image. Compared with
the AEM and the PFM, the TDA can extract 1D spectra with both higher SNR and
resolution. Theoretically, our algorithm discussed in this
paper can be applied to a 2D image of any size and PSFs
with any shape, even there is serious cross talk
between fibers.

We further consider how to set  precision parameter $\epsilon$  to balance the computation time against the result precision,
    and find that setting the calculation precision similar to the noise level should be a reasonable choice. Because our method converges very fast, for images with not so extreme
   cross talk, the computation time does not depend so much on fiber distance, usually in 2 to 3 iterations, the result will be around the peak SNR.
     For our simulation with 250 fibers and 4k pixels in wavelength direction, the computation times are about 2 hours for the images with fiber distance greater than 6 pixels and 3.5 hours for 6 pixels.
 Moreover, our algorithm could be easily transferred to a parallel algorithm, and the
computation speed will be boosted. 
\par
In summary, compared with the traditional PFM and AEM, deconvolution
method is the most consistent with physical process that  2D
spectra are recorded by the CCD, so it can extract the most accurate fluxes from a 2D
image, especially for multi-fiber spectra. Beside these, spectra
extracted by our deconvolution method have higher resolution and SNR,
so it is the most promising extraction method.
\par
However, there are still some uncertainties on weak features
around strong emission lines in the 1D spectra extracted by the TDA.
Laborious efforts are still needed to solve the problem of
noise influence. Besides the TDA, there are many other methods
in the literature to attenuate the noise influence in signal
processing, such as  the linear regularized methods, the Bayesian
methodology and the wavelet-based deconvolution methods
(see \cite{sta02}). To study how to attenuate the noise
influence on the deconvolution method will be the topic
of our future work.
\par

\acknowledgments
This research is supported by the Natural Science Foundation
of China for the Youth under grants Y011161001 and NSFC Key
Program NSFC-11333004. We also thank the referee for valuable
advices. Guoshoujing Telescope (the Large Sky Area Multi-Object
Fiber Spectroscopic Telescope LAMOST) is a National Major
Scientific Project built by the Chinese Academy of Sciences.
Funding for the project has been provided by the National
Development and Reform Commission. LAMOST is operated and
managed by the National Astronomical Observatories, Chinese
Academy of Sciences.

\begin{landscape}
  \begin{table}[p]

\caption{Computation time and calculation precision in each iteration for images with different fiber distances}
\label{tab:cputime}
\tiny
\begin{tabular}{|c||c|c|c|c||c|c|c|c||c|c|c|c||c|c|c|c|} \hline
{}& \multicolumn{4}{c||}{Computation time(s) }&\multicolumn{4}{c||}{Precision (difference of RES, see section\ref{sec:44}))}& \multicolumn{4}{c||}{SNR(Brihgt Fiber) }&\multicolumn{4}{c|}{SNR(Faint Fiber) }\\ \hline
 { \diagbox[height=35pt]{\tiny{Iteration}}{\raisebox{0.5ex}[0pt]{\tiny{fiber}}\\\raisebox{1.5ex}[0pt]{\tiny{distance}}}}& 15&  10 & 8 & 6 &15& 10 & 8 & 6 &15& 10 & 8 & 6 &15& 10 & 8 & 6 \\ \hline
 1	 &	2487  & 2467 & 2520   & 2472 & 2.35e13 & 2.35e13 & 2.35e13 & 2.30e13 & 3.58&3.55 &3.57 & 3.62 & 4.57 &1.38 &0.17 & 0.04 \\ \hline
 2	 &	5035  & 5015 & 5030  & 5032 & 2.95e9 & 3.50e9 & 3.58e10 & 6.28e11 & 15.07 &14.74 & 15.04 &14.16 & 4.58 & 4.74 & 4.19&0.80 \\ \hline
 3	 &	7550  & 7507 & 7557  & 7535 & 2.12e7 & 2.13e7 & 7.89e7 & 2.01e10 &16.34 &16.18 &16.70 &16.06 & 4.58&4.71 &4.71 &4.16 \\ \hline
 4	 &	\textbf{10057}  & \textbf{9997} & \textbf{10085}  & 10027  & \textbf{2.16e5} & \textbf{2.62e5} & \textbf{3.65e5} & 6.89e8   & 16.27 &16.14 & 16.68 &16.02 & 4.58& 4.71 &4.70  &3.98 \\ \hline
 5	 &	12555  & 12490 & 12602  & 12575 & 4.27e3 &  4.37e3 & 4.50e3 & 2.39e7 & 16.25 & 16.13 &16.67 & 16.00 & 4.58 & 4.71 & 4.70 & 4.27 \\ \hline
 6	 &	15057  & 14977 &15087   & \textbf{15120} & 84.9 &   84.9 & 87.2 &\textbf{ 9.55e5}  & 16.25 & 16.13 & 16.67 & 16.00 & 4.58 & 4.71 & 4.70 & 4.30\\ \hline
 7	 &	17552  & 17470 & 17600   & 17662 & 1.94 & 1.93 & 1.99 & 4.15e4  & 16.25 & 16.13 &16.67 &16.00 & 4.58 &4.71 &4.70 & 4.31 \\ \hline
8	 &	20097   & 20017 & 20092   & 20167 & 4.90e-2 & 4.89e-2 & 5.05e-2 & 1.86e3  & 16.25 & 16.13 & 16.67 & 16.00 & 4.58 & 4.71 & 4.70 &4.31 \\ \hline
9	 &	22592   & 22570 & 22590    & 22707  & 1.35e-3 & 1.35e-3 & 1.40e-3 & 7.75  & 16.25 & 16.13 & 16.67 & 16.00 & 4.58 & 4.71 & 4.70 &4.31 \\ \hline
\end{tabular}
\normalsize
\end{table}
\end{landscape}


\begin{thebibliography}{}



%
%
%
%
%
\bibitem[Ballester et al.(2000)]{bal00}
 Ballester, P., Modigliani, A., Boitquin, O., et al.,
    2000, The Messenger, 101, 31
\bibitem[Bolton \& Schlegel(2010)]{bol10} Bolton, A. S. \& Schlegel,
    D. J., 2010, \pasp, 122, 248
\bibitem[Cui et al.(2012)]{cui12} Cui, X. Q., Zhao, Y. H., Chu, Y. Q., et al., 2012, RAA, 12, 1197
\bibitem[Davies \& Kasper(2012)]{dav12} Davies, R. \& Kasper, M.,
    2012, ARA\&A, 50, 305
\bibitem[Horne(1986)]{hor86} Horne, K., 1986, \pasp, 98, 609
\bibitem[Hynes(2002)]{hyn02} Hynes, R. I., 2002, A\&A, 382, 752
\bibitem[Kelz et al.(2006)]{kel06} Kelz, A., Verheijen, M. A. W., Roth, M. M., et al. 2006, \pasp, 118, 129

\bibitem[Lebouteiller et al.(2010)]{leb10} Lebouteiller, V., Bernard-Salas, J., Sloan, G. C., et al.,
    2010, \pasp, 122, 231
\bibitem[Marsh(1989)]{mar89} Marsh, T. R., 1989, \pasp, 101, 1032
\bibitem[Mukai(1990)]{muk90} Mukai, K., 1990, \pasp, 102, 183
\bibitem[Piskunov \& Valenti(2002)]{pis02} Piskunov, N. E. \& Valenti,
    J. A., 2002, A\&A, 385, 1095
\bibitem[Puetter et al.(2005)]{pue05} Puetter, R. C., Gosnell, T. R.,
     Yahil, A., 2005, ARA\&A, 43, 139

\bibitem[Robertson(1986)]{rob86} Robertson, J. G., 1986, \pasp, 98, 1220
\bibitem[S\'{a}nchez(2006)]{san06} S\'{a}nchez, S. F., 2006,
    Astronomische Nachrichten, 327, 850
\bibitem[Sharp \& Birchall(2010)]{sha10} Sharp, R. \& Birchall,
    M. N., 2010, \pasa, 27, 91
\bibitem[Starck et al.(2002)]{sta02} Starck, J. L., Pantin, E.,  Murtagh, F.,
    2002, \pasp, 114, 1051

\bibitem[Tikhonov(1963)]{tik63} Tikhonov, A. N., 1963,
    Soviet Mathematics Doklady, 4, 1624
\bibitem[Tikhonov \& Arsenin(1977)]{tik77}  Tikhonov, A. N. \&
    Arsenin, V. Y., 1977, Solutions of Ill-posed Problems, New York: Wiley

\bibitem[Verschueren \& Hensberge(1990)]{ver90} Verschueren, W.
    \& Hensberge, H., 1990, A\&A, 240, 216
\bibitem[Hanuschik (2003)]{skyemi} Hanuschik, R. W., 2003, A\&A, 407, 1157

\end{thebibliography}
\end{document}